\documentclass[twocolumn,tighten]{aastex62}
\usepackage{amssymb, amsmath,framed}

\usepackage{bm}
\expandafter\ifx\csname package@font\endcsname\relax\else
 \expandafter\expandafter
 \expandafter\usepackage
 \expandafter\expandafter
 \expandafter{\csname package@font\endcsname}
\fi
\def\bq{\begin{equation}}
\def\eq{\end{equation}}
\def\bqy{\begin{eqnarray}}
\def\eqy{\end{eqnarray}}
\begin{document}
\title{\large{Brown Dwarf Atmospheres As The Potentially Most Detectable And Abundant Sites For Life}}

\correspondingauthor{Manasvi Lingam}
\email{manasvi.lingam@cfa.harvard.edu}

\author{Manasvi Lingam}
\affiliation{Institute for Theory and Computation, Harvard University, Cambridge MA 02138, USA}
\affiliation{Department of Aerospace, Physics and Space Sciences, Florida Institute of Technology, Melbourne FL 32901, USA}

\author{Abraham Loeb}
\affiliation{Institute for Theory and Computation, Harvard University, Cambridge MA 02138, USA}

\begin{abstract}
We show that the total habitable volume in the atmospheres of cool brown dwarfs with effective temperatures of $\sim 250$-$350$ K is possibly larger by two orders of magnitude than that of Earth-like planets. We also study the role of aerosols, nutrients and photosynthesis in facilitating life in brown dwarf atmospheres. Our predictions might be testable through searches for spectral edges in the near-infrared and chemical disequilibrium in the atmospheres of nearby brown dwarfs that are either free-floating or within several AU of stars. For the latter category, we find that the James Webb Space Telescope (JWST) may be able to achieve a signal-to-noise ratio of $\sim 5$ after a few hours of integration time per source for the detection of biogenic spectral features in $\sim 10^3$ cool brown dwarfs.\\
\end{abstract}

\section{Introduction} \label{SecIntro}
Whenever environments conducive to the origin and sustenance of life (i.e., habitable environments) are studied, they are almost invariably assumed to be based either on the surface or beneath the surface. The Earth is a classic example of the former, whereas the subsurface oceans of Europa and Enceladus constitute well-known candidates for the latter \citep{Shap67,NP16}. However, there is one crucial environment that is often neglected in considerations of habitability, namely, the atmosphere \citep{SMI08,PD14}. The absence of a continuous solid substrate and reduced protection against cosmic rays as well as ultraviolet (UV) radiation constitute some of the reasons commonly advanced to justify why environmental conditions are not suitable for the presence of aerial biospheres. 

However, at an appropriate altitude, favorable physicochemical conditions may exist for life, including liquid water, moderate temperatures and nutrients. Several studies have proposed that the cloud layer of Venus at $\sim 50$ km is potentially habitable in the above respects, even though its surface is far too hot for life-as-we-know-it. This proposal dates back to more than $50$ years ago \citep{MS67} (see also \citealt{SL70}) and has gained some traction from the 1990s onwards \citep{Grin97,Cock99,SGA04,DNP,LMS18}. 

Apart from Venus, a few early studies also examined the possibility of life in the atmospheres of Jovian planets \citep{SoRe67,Shap67,SS76}. These papers were complemented by laboratory experiments mimicking Jovian atmosphere that yielded a number of valuable prebiotic compounds such as amino and imino acids (after acid hydrolysis), aminonitriles, formaldehyde and hydrogen cyanide \citep{Sag60,PM73,Pon76,SM87}; numerical simulations also indicate that some of the above molecules as well as considerable amounts of small hydrocarbons could exist in the upper atmospheres of substellar objects \citep{BRH13,SHD14}.

The dividing line between giant planets and brown dwarfs is not a very sharply delineated one, as there has been much debate regarding the classification of objects with masses of $\sim 3$-$10\,M_J$, where $M_J$ denotes the mass of Jupiter, and with effective temperatures $< 500$ K; these objects do not appear to possess the capacity for deuterium fusion \citep{Cab18}. As some of the salient physical and chemical characteristics are similar across giant planets and cool brown dwarfs \citep{BHL01,HBL02,HC14,Bai14,MR15}, it is natural to inquire whether the atmospheres of brown dwarfs are conducive to the origin and sustenance of life.

To the best of our knowledge, the only paper that has addressed the issue of atmospheric habitability for cool brown dwarfs (of spectral class Y) is \citet{YPB17}.\footnote{However, an early qualitative exploration of potential lifeforms in the atmospheres of brown dwarfs and gas giants can be found in the imaginative, but sadly forgotten, book by \citet{Shap67}.} By drawing upon an organism lifecycle model, \citet{YPB17} found that putative microbes up to an order of magnitude larger, and masses a few orders of magnitude higher, than typical microbes on Earth could be supported in the presence of atmospheric convection. In that paper, the number of free-floatng cool Y dwarfs in the Miky Way was also investigated, and an estimate of $\sim 10^9$ was derived based on observational constraints. Our work will deal with other aspects of brown dwarf atmospheric habitability and putative biospheres that were not explicitly tackled in \citet{YPB17}.

The outline of our paper is as follows. In Section \ref{SecMaxHZ}, we estimate the maximum habitable volume that is encapsulated by the atmospheres of cool brown dwarfs. We continue by exploring the characteristics of putative aerial biospheres in Section \ref{SecLifeBD} with a focus on prebiotic chemistry, abiogenesis, and nutrient and energy supplies. Subsequently, we identify possible biosignatures that might arise in these atmospheres and the methods of detecting them in Section \ref{SecLifeD}. Finally, we summarize our central results in Section \ref{SecConc}.

\section{Assessing the maximum habitable volume}\label{SecMaxHZ}
We shall start by presenting heuristic estimates for the total potentially habitable volume encompasses by two different classes of objects. Strictly speaking, our analysis yields a potential upper bound for the habitable volume if habitability is narrowly interpreted as conditions permitting the existence of liquid water. Although we do not explicitly employ the word ``potential'' henceforth, it should be understood that we are dealing with: (i) potential habitability and, (ii) a potential upper bound on the habitable volume. The ``volume'' under question is four-dimensional as it encompasses not only the spatial coverage but also the temporal duration of habitability. 

For a single object, we denote the spatial habitable volume by $V$ and the habitability interval by $\tau$, yielding the four-dimensional (4D) volume $(\mathcal{V}$) defined as $\mathcal{V} \sim V \tau$. However, this applies only to a single object. In order to determine the total 4D volume spanned by a given class of objects, we introduce the function
\begin{equation}\label{GamDef}
    \Gamma = \int \mathcal{V}\,dN = \int V \tau\,dN,
\end{equation}
where $dN$ is the number of objects within a particular interval, as we shall describe hereafter. One of the chief points worth appreciating is that $V$, $\tau$ and $dN$ are functions of the object's mass.

\subsection{Atmospheric habitable zone in brown dwarfs}\label{SSecAHZBD}
The first class of systems we evaluate are the the atmospheric habitable regions of brown dwarfs. We use the subscript `BD' subsequently to identify this category. 

We begin by evaluating $\tau_\mathrm{BD}$. This is found by calculating the duration of time over which the brown dwarf has an effective temperature of $\sim 250$-$350$ K. There are three reasons behind our choice of this thermal range: (a) a number of interesting molecular species from the perspective of prebiotic chemistry and biochemistry may be present, (b) the existence of aerosols is feasible, and (c) atmospheric water vapor and clouds comprising volatile ices are believed to exist \citep{MMF14,SMA16,MSA18,Hel19}.\footnote{It must however be noted that water vapor can exist in the atmospheres of bodies with higher effective temperatures; for instance, water vapor lines have been detected in Arcturus \citep{RLR02}, whose surface temperature is $4290$ K.} Another advantage of focusing on the above temperature range is that the upper atmospheres of such brown dwarfs are anticipated to share some resemblance with the lower atmosphere of Earth \citep{YPB17}.

In order to calculate $\tau_\mathrm{BD}$ in our simple model, we will make use of the analytical expression derived in \citet{BL93}. Following equation (2.58) of \citet{BL93}, the effective temperature ($T_\mathrm{eff}$) of the brown dwarf is given by
\begin{eqnarray}\label{Teffrel}
 && T_\mathrm{eff} \approx 1551\,\mathrm{K}\,\left(\frac{t_\mathrm{BD}}{1\,\mathrm{Gyr}}\right)^{-0.324} \left(\frac{M_\mathrm{BD}}{0.05\,M_\odot}\right)^{0.827} \nonumber \\
 && \hspace{0.3in} \times \left(\frac{\kappa_R}{0.01\,\mathrm{cm^2/g}}\right)^{0.088},  
\end{eqnarray}
where $t_\mathrm{BD}$ and $M_\mathrm{BD}$ are the age and mass of the brown dwarf, respectively, and $\kappa_R$ denotes the Rosseland mean opacity of the brown dwarf close to its photosphere. The value of $\kappa_R$ has a complicated dependence on the wavelength, density, temperature and the chemical species present \citep{FLF14}; the weighted mixing ratio for cool brown dwarfs at $T_\mathrm{eff} = 250$ K has been provided in Figure 4 of \citet{MSA18}. Owing to the very weak dependence on $\kappa_R$ in (\ref{Teffrel}), this factor may be set aside in our order-of-magnitude analysis. 

After rearranging the preceding expression and solving for $\tau_\mathrm{BD}$,\footnote{We opt to normalize $M_\mathrm{BD}$ in units of $M_\odot$ because we will deal with stellar habitable zones subsequently. However, in the later sections, we will revert to the conventional normalization factor of Jupiter's mass ($M_J$).} we have
\begin{equation}\label{TauBD}
    \tau_\mathrm{BD} \approx 3.8 \times 10^{5}\,\mathrm{Gyr}\, \left(\frac{M_\mathrm{BD}}{M_\odot}\right)^{2.55},
\end{equation}
The mass range for brown dwarfs lacks precise upper and lower bounds due to ambiguities stemming from their definition. Based on the orthodox definition entailing deuterium burning, the range $0.01 \lesssim M_\mathrm{BD}/M_\odot \lesssim 0.06$ is often employed \citep{Cab18}. However, this ignores the existence of overmassive brown dwarfs \citep{FL19} as well as sub-brown dwarfs such as WISE J085510.83-071442.5 whose mass is $< 0.01 M_\odot$ \citep{Luh14}. In this paper, we will employ the range $0.005 < M_\mathrm{BD}/M_\odot < 0.07$, which is mostly coincident with the conventional limits delineated earlier.

Next, we consider the volume $V_\mathrm{BD}$ encompassed by the atmospheric habitable zone. As the atmosphere constitutes a spherical shell, we have
\begin{equation}
   V_\mathrm{BD} \approx 4 \pi R_\mathrm{BD}^2 \mathcal{H}_\mathrm{BD}, 
\end{equation}
with $R_\mathrm{BD}$ being the radius of the brown dwarf that is given by equation (2.37) of \citet{BL93}:
\begin{equation}\label{RaBD}
 R_\mathrm{BD} \approx 3.5 R_\oplus\,\left(\frac{M_\mathrm{BD}}{M_\odot}\right)^{-1/3}, 
\end{equation}
where we have opted to normalize $R_\mathrm{BD}$ in units of $R_\oplus$ because we will subsequently study Earth-sized planets in the habitable zones of stars. Next, we must assess the characteristic value of $\mathcal{H}_\mathrm{BD}$, the vertical layer in which habitable conditions exist. Given that extremophiles on Earth have been documented at temperatures ranging from $T_\mathrm{min} \sim 258$ K to $T_\mathrm{max} \sim 395$ K \citep{Cla14,McK14}, these limits can be utilized to determine $\mathcal{H}_\mathrm{BD}$. Of course, in using these limits, we are working with the implicit assumption that extraterrestrial life has the same thermal limits as organisms on Earth. While this premise is indubitably geocentric, it is plausible that the thermal range for extraterrestrial life, as set by generic physicochemical constraints, is not far removed from $T_\mathrm{min}$ and $T_\mathrm{max}$ \citep{BXY15,VH18}. A subtle, but important, distinction is worth pointing out here: $T_\mathrm{min}$ and $T_\mathrm{max}$ refer to the local \emph{atmospheric} temperature in the habitable zone, whereas the range for $T_\mathrm{eff}$ introduced at the beginning of Section \ref{SSecAHZBD} corresponds to the blackbody temperature of the brown dwarf as it cools.  

From inspecting the pressure-temperature diagram presented in Fig. 6 of \citet{MMF14} for cool Y dwarfs, it is found that the pressure varies by a factor of $\sim 10$ across this temperature range; in quantitative terms, the pressure in this region is typically $\sim 0.1$-$1$ bar. This pressure range is compatible with the pressure range tolerated by Earth-based organisms \citep{McK14,DD18}. For an isothermal atmosphere in hydrostatic equilibrium, the above result for the pressure implies that $\mathcal{H}_\mathrm{BD}$ would be a few times the scale height of the brown dwarf. As the scale height is inversely proportional to the surface gravitational acceleration $g$, we specify
\begin{equation}
 \mathcal{H}_\mathrm{BD} \sim 2.4 \times 10^{-2}\,\mathrm{km}\,  \left(\frac{M_\mathrm{BD}}{M_\odot}\right)^{-5/3} 
\end{equation}
after making use of equation (2.51) from \citet{BL93} with the normalization obtained from Section 2.1 of \citet{YPB17}.

The last term that we need to tackle is $dN_\mathrm{BD}$, which can be simplified to $(dN/dM_\mathrm{BD})\,dM_\mathrm{BD}$. In other words, we wish to determine $\xi_\mathrm{BD} \equiv dN/dM_\mathrm{BD}$, but this remains difficult to estimate, as the abundance of brown dwarfs is not rigorously constrained. We will make use of the initial mass function described in \citet{TK07} and \citet{KWP13} that is in reasonable agreement with numerical simulations and empirical observations \citep{TPK15}; it has the form
\begin{equation}\label{SSBDIMF}
  \xi_\mathrm{BD} \approx 9 \times 10^{-2}\,\mathcal{C} \left(\frac{M_\mathrm{BD}}{M_\odot}\right)^{-0.3}, 
\end{equation}
where $\mathcal{C}$ represents a normalization constant that drops out of our subsequent analysis. However, for late-type brown dwarfs (spectral classes T and Y), the mass function is better described by a power-law exponent of $-0.6$ \citep{KMS19}; using this scaling changes our subsequent results by a factor of $\sim 3$. Therefore, after substituting all of the preceding relations into (\ref{GamDef}), we can determine $\Gamma_\mathrm{BD}$. 

Before moving ahead, a comment on putative organisms in atmospheric habitable zones of brown dwarfs is in order. On the one hand, due to the gravitational settling, the microbes drift downward at the terminal velocity. On the other, convective updraft will act to transport microbes in the opposite direction. \citet{YPB17} developed numerical and analytical models taking into account both of these factors and found that organisms up to $\sim 10$ times larger than typical Earth-based microbes (with sizes of $\sim 1$ $\mu$m) might exist as stable populations in the atmospheric habitable zones of cool Y dwarfs. \citet{YPB17} found that strong convection (i.e., higher velocities) favors the evolution of larger organisms, and vice-versa.

\subsection{Circumstellar habitable zones around main-sequence stars}
The next category of systems we consider are Earth-sized worlds in the habitable zones (HZs) of main-sequence stars \citep{Dole64,KWR93,KRK13,Ram18}. As mentioned earlier, we are interested only in potential habitability, owing to which we consider all worlds that are Earth-sized situated in the HZ, regardless of whether they actually host liquid water or not. We will use the subscript `$\star$' to indicate that we are studying worlds in the HZs of stars.

The first term of interest to us is $\tau_\star$, namely, the period over which the planet resides in the HZ. In theory, the upper bound is determined by the stellar lifetime, but the actual value is lower because the planet will experience a greenhouse effect and possibly end up being desiccated \citep{CK92,GW12}. The duration of temporal habitability was investigated by \citet{RCO13} for planets receiving Earth-like insolation. The results were subsequently parametrized as simple scaling relations by \citet{LL19}, which we will adopt herein as follows:
\begin{eqnarray}\label{tHZ}
&& t_{\star} \sim 5.5\,\mathrm{Gyr}\,\left(\frac{M_\star}{M_\odot}\right)^{-2} \quad \quad M_\star > M_\odot, \nonumber \\
&& t_{\star} \sim 5.5\,\mathrm{Gyr}\,\left(\frac{M_\star}{M_\odot}\right)^{-1} \quad \quad\, 0.5 M_\odot < M_\star < M_\odot, \nonumber \\
&& t_{\star} \sim 4.6\,\mathrm{Gyr}\,\left(\frac{M_\star}{M_\odot}\right)^{-1.25} \quad M_\star < 0.5 M_\odot,
\end{eqnarray}
where $M_\star$ is the mass of the host star. Note that $t_{\star}$ may represent an upper bound on the habitability lifetime of M-dwarf exoplanets. Such planets are subject to intense stellar winds \citep{LL17,Lin18,DLM17,DJL18,DHL19} and elevated X-ray and extreme UV radiation fluxes \citep{LB15,BSO17}, and could therefore be depleted of their atmospheres and water over sub-Gyr timescales; for a recent review of this subject, see \citet{Man19}.

The habitable volume for an ``Earth-like'' planet in the HZ is found via
\begin{equation}
   V_{\star} \approx 4 \pi R_\oplus^2 \mathcal{H}_\oplus, 
\end{equation}
where the habitable ``height'' $\mathcal{H}_\oplus$ needs to be determined. Habitability is modelled as being restricted to the regions of the planet where the temperatures are between $T_\mathrm{min}$ and $T_\mathrm{max}$. This includes not only the surface of the planet but also its atmosphere and lithosphere. As we are dealing with worlds that resemble Earth in terms of their geochemical characteristics, we are free to use Earth as a proxy. First, let us consider the atmospheric habitable zone of the Earth. Based on the thermal limits and the atmospheric temperature profile \citep{Jac99}, we find that $\sim 5$ km in the troposphere, $\sim 10$ km near the stratopause, and $\sim 10$ km in the thermosphere meet this criterion. Next, if we choose an average geothermal gradient of $\sim 2.5 \times 10^{-2}$ K/m, we find that the habitable depth is $\sim 4$ km by starting with a surface temperature of $\sim 290$ K. Thus, after adding up these contributions, we end up with $\mathcal{H}_\oplus \sim 29$ km.  

The last quantity to determine is $\xi_\star \equiv dN/dM_\star$ since $dN_\star = \eta_\oplus \xi_\star\,dM_\star$. The additional factor $\eta_\oplus$ accounts for the number of Earth-sized planets in the HZs of their host stars. The estimated value of $\eta_\oplus$ is uncertain by nearly an order of magnitude \citep{Kal17}. We will treat $\eta_\oplus$ as being roughly independent of $M_\star$, and will adopt the conservative choice of $\eta_\oplus \sim 0.1$, even though recent studies indicate that $\eta_\oplus \lesssim 0.35$ for K- and G-type stars \citep{ZH19}. To maintain consistency with the previous section, we adopt the same initial mass function \citep{TK07,TPK15}, thus yielding
\begin{eqnarray}
&& \xi_\star \approx 3.2 \times 10^{-2}\,\mathcal{C} \left(\frac{M_\star}{M_\odot}\right)^{-1.3} \,\,\, 0.07 M_\odot < M_\star < 0.5 M_\odot, \nonumber \\
&& \xi_\star \approx 1.6 \times 10^{-2}\,\mathcal{C} \left(\frac{M_\star}{M_\odot}\right)^{-2.3} \,\,\, M_\star > 0.5 M_\odot.
\end{eqnarray}
With all of the factors assembled, we are now in a position to calculate $\Gamma_\star$ after drawing upon (\ref{GamDef}). The lower bound for the stellar mass is chosen to be $0.07 M_\odot$ whereas the upper bound is $2 M_\odot$; increasing the upper bound to infinity alters our results by $< 20\%$.

\subsection{Ratio of habitable volumes}
The chief quantity of interest is the following ratio:
\begin{equation}
    \Delta_\mathcal{V} \equiv \frac{\Gamma_\mathrm{BD}}{\Gamma_\star}.
\end{equation}
Clearly,  $\Delta_\mathcal{V} \gg 1$ implies that most of the habitable volume is concentrated in the atmospheres of brown dwarfs relative to that encompassed by Earth-like planets in the habitable zone and vice-versa. After simplifying $\Gamma_\mathrm{BD}$ and $\Gamma_\star$ by utilizing the results from the preceding sections, we arrive at $\Delta_\mathcal{V} \sim 52$. Therefore, this result that implies the maximum potentially habitable volume is in the atmospheres of brown dwarfs and not Earth-like planets in HZs of stars.

At this stage, we note that life around stars is not constrained to exist only in the HZ. As the examples of Europa and Enceladus illustrate, life may also exist in subsurface oceans underneath icy envelopes. Estimating the ratio of the four-dimensional volumes for Earth-like and subsurface worlds is a much more challenging endeavor, owing to which we shall not address this question in detail. However, the following points should be borne in mind with regards to this matter.
\begin{itemize}
    \item It has been estimated that the maximal number of worlds with subsurface oceans that are larger than Europa is $\gtrsim 10^3$ times higher than Earth-like planets in the HZ \citep{LiLo19}.
    \item The volume of oceans in such subsurface ocean worlds might be a few times higher with respect to Earth depending on their water inventory; Europa, in particular, may possess $\sim 2.5$ times more water than Earth provided that its ocean depth is $\sim 100$ km \citep{Chy00,Lun17}.
    \item Even in the absence of tidal heating, subsurface ocean worlds of the type specified above could retain their oceans over Gyr timescales via radiogenic heating \citep{SS03}.
\end{itemize}
Based on these considerations, it is conceivable that the total habitable volume spanned by subsurface ocean worlds is a couple of orders of magnitude higher than Earth-like planets in the HZ. In this event, the total volume spanned by this category would be comparable to that encompassed by the atmospheres of brown dwarfs.

\section{The prospects for life in brown dwarf atmospheres}\label{SecLifeBD}
Hitherto, we have restricted ourselves to the potentially habitable volume covered by brown dwarf atmospheres. However, a well-known fact is that habitability requires more than just the appropriate thermal limits and the existence of liquid water \citep{Lam09,SBJ16,Man19}. We will thus explore certain aspects pertaining to habitability and putative life in the atmospheres of cool brown dwarfs. In most instances, as we possess no knowledge whatsoever of the putative organisms that can inhabit brown dwarf atmospheres, \emph{faute de mieux}, we draw upon analogs from Earth and adapt them accordingly.

\subsection{Biomass in atmospheric habitable zones}\label{SSecBMAHZ}
Next, we estimate the \emph{upper bound} on the possible biomass in brown dwarf atmospheres. Naturally, the lower bound is trivially zero in the event that life is ruled out altogether in aerial settings. While we are not aware of any organisms that complete their entire life cycles in Earth's atmosphere, comparatively few studies have been undertaken and additional research is necessary \citep{Smi13}. Although permanent ecosystems in Earth's stratosphere are ``unlikely'' due to the harsh environment \citep{SGS10}, the conditions in its troposphere are less extreme. Moreover, for reasons documented a few paragraphs below, the atmospheric habitable zones of cool brown dwarfs might be more clement than the Earth's stratosphere in some respects. Finally, the possibility of these habitable zones being seeded by microbes from elsewhere ought not be discounted altogether as per some hypotheses \citep{WAW06}.

In order to estimate the biomass in brown dwarf atmosphere, we will utilize data derived from the Earth's atmosphere. In doing so, we are implicitly invoking a strong version of the Copernican Principle, whose validity is by no means confirmed. A more realistic treatment would attempt to derive the biomass based on nutrient (and energy) availability, but carrying out this calculation requires an in-depth knowledge of the limiting nutrients and their coupled biogeochemical cycles. Hence, we opt to use empirical results from the Earth, owing to which the ensuing findings should be viewed with appropriate caution.

We denote the characteristic number density of microbes in the atmosphere by $\rho_m$, thus yielding a total biomass ($M_\mathrm{bio}$) of
\begin{equation}\label{BioMassT}
    M_\mathrm{bio} \sim 4\pi \rho_m R_\mathrm{BD}^2 \mathcal{H}_\mathrm{BD}.
\end{equation}
As per Table 1 of \citet{FNK16}, the average global mass density of microbes is $\rho_m \sim 10^{-10}$ kg/m$^3$. However, in localized regions, the concentration of airborne microbes can be as high as $\sim 10^{11}$ m$^{-3}$ \citep{APS07}, which translates to a mass density of $\rho_m \sim 10^{-4}$ kg/m$^3$ after presuming that the mass of a single microbe is $\sim 10^{-15}$ kg. 

Both of the above estimates pertain to modern-day Earth. Instead, let us turn our attention to Archean Earth, which may have possessed a thick haze cover. From Figure 1 of \citet{ADG16}, we find that the global aerosol density ranges from $\sim 10^{5}$ m$^{-3}$ in the troposphere to $\sim 10^{10}$ m$^{-3}$ in the thermosphere; the particle radius of the aerosols is a few times smaller than $1$ $\mu$m. On modern Earth, around $20$-$30\%$ of all aerosols with sizes $> 0.5$ $\mu$m have been documented to host microbes \citep{BMH12,FNK16}. Hence, if we adopt a similar fraction of $25\%$ for Archean Earth, we find that the biomass density is characterized by a range of $\rho_m \sim 2.5 \times 10^{-11}$ kg/m$^3$ to $\rho_m \sim 2.5 \times 10^{-6}$ kg/m$^3$. Note that we have neglected factors other than the aerosol density to estimate the atmospheric biomass density. 

In actuality, other factors such as the access to nutrients and thermodynamic disequilibrium as well as environmental parameters will play a major role in determining atmospheric biomass density. However, it is worth appreciating that certain microbes on Earth can survive in the stratosphere despite the presence of considerable desiccation, cold temperatures, nutrient deficiency, and high ultraviolet and ionizing radiation fluxes \citep{HGT13,DD18}. In contrast, the atmospheric habitable zones might have access to water and nutrients (the latter is described later), receive lower doses of radiation due to clouds and hazes \citep{MMF14,Hel19}, and (by definition) be characterized by temperatures within the thermal limits of Earth-based life. 

In the clouds of Venus, theoretical calculations by \citet{LMS18} suggest that the maximum biomass loading for aerosols with sizes $> 2$ $\mu$m is $\rho_m \sim 1.4 \times 10^{-5}$ kg/m$^3$. As Jupiter is the closest analog that we possess for a brown dwarf in our Solar system, it is worth examining the Jovian atmosphere in more detail. The nephelometer carried by the \emph{Galileo} spacecraft detected particles throughout its descent from $\sim 0.5$ to $\sim 12$ bars. The mean particle radius ranged between $\sim 0.5$-$4.0$ $\mu$m and the observed number density spanned $\sim 1.4 \times 10^5$ to $3.4 \times 10^7$ m$^{-3}$ \citep{WBF04}; the maximum number density was detected at a pressure of $0.75$-$1.35$ bar, which is close to the Earth's surface pressure. Saturn also has aerosols in its atmosphere, whose characteristics are broadly similar to Jupiter \citep{WBK09}. 

Finally, we note that Titan's atmosphere comprises a thick organic haze that is primarily produced by photochemistry \citep{CHH12,Hor17}. Constraints derived from the Descent Imager/Spectral Radiometer onboard the \emph{Cassini-Huygens} probe apparently indicate that the number density of aerosols at an altitude of $80$ km is $\sim 5 \times 10^6$ m$^{-3}$ with a possible mean particle size of $\sim 0.7$-$2$ $\mu$m \citep{TDE08}. Photochemical and electrical discharge experiments conducted to simulate Titan's atmosphere have yielded higher aerosol densities of $4.6 \times 10^9$ to $3 \times 10^{12}$ m$^{-3}$, albeit at smaller sizes ($< 1$ $\mu$m), as seen from Table 1 of \citet{HT13}.

\begin{figure}
\includegraphics[width=7.5cm]{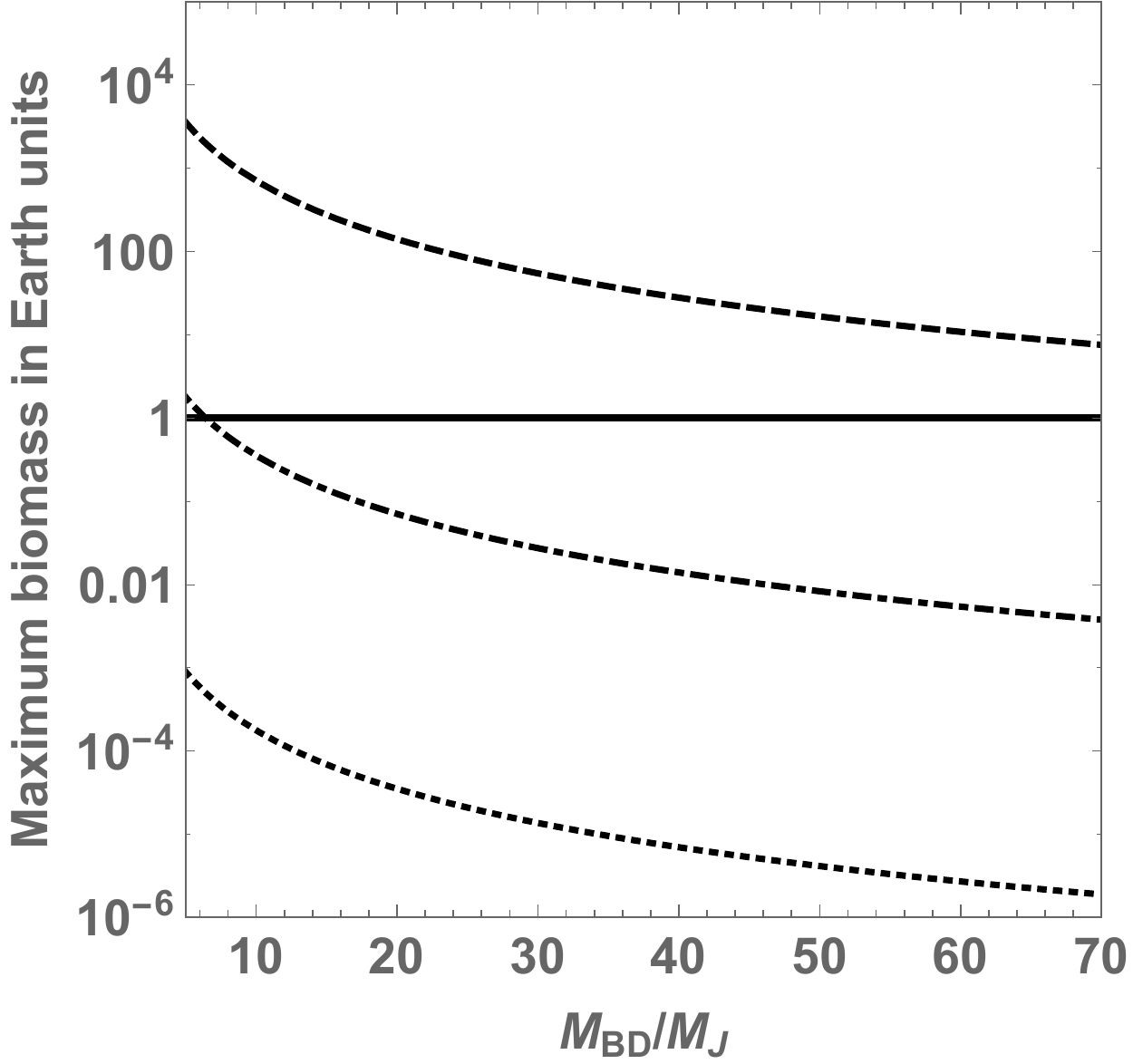} \\
\caption{The maximum biomass that may exist in cool brown dwarf atmospheres (normalized by Earth's biomass) as a function of the brown dwarf mass ($M_\mathrm{BD}$) in units of Jupiter's mass ($M_J$). The dashed, dotted-dashed and dotted curves correspond to the biomass obtained by using a biomass density of $\rho_\mathrm{max}$, $\bar{\rho}$ and $\rho_\mathrm{min}$, respectively. The solid horizontal line corresponds to the case where the atmospheric biomass is equal to that of Earth.}
\label{FigBiomass}
\end{figure}

Determining the aerosol density of brown dwarfs is not easy as it depends upon the chemical species present as well as the spectral type and altitude. Under the assumptions of hydrostatic equilibrium and complete condensation, the upper bound on the column mass density of the particles ($\sigma_c$) is \citep{MR15}:
\begin{equation}
  \sigma_c \approx f \left(\frac{m_c}{\bar{m}}\right)\left(\frac{P_c}{g}\right),  
\end{equation}
where $m_c$ is the mass of the condensate particle, $\bar{m}$ represents the mean molecular weight of the atmosphere, $f$ denotes the mixing ratio for the condensible species, and $P_c$ is the pressure at which condensation occurs. However, note that this formula only yields the column density and not the number density. 

If we assign the same values of the maximum and minimum biomass density to brown dwarfs by drawing upon the prior examples from our Solar system, we find that the lower bound is $\rho_\mathrm{min} \sim 2.5 \times 10^{-11}$ kg/m$^3$ whereas the upper bound is $\rho_\mathrm{max} \sim 10^{-4}$ kg/m$^3$. By computing the geometric mean of these two quantities, we end up with $\bar{\rho} \sim 5 \times 10^{-8}$ kg/m$^3$. We are now in a position to calculate (\ref{BioMassT}) for these choices of the biomass density. However, it is more instructive to estimate the total biomass normalized to that of the Earth, i.e., we introduce the ratio
\begin{equation}\label{RelBioM}
    \delta_\mathrm{bio} \equiv \frac{M_\mathrm{bio}}{M_\mathrm{bio,\oplus}}.
\end{equation}
In determining $M_\mathrm{bio,\oplus}$, we observe that the total amount of biogenic carbon on our planet is $5.5 \times 10^{14}$ kg \citep{BPM18}. In order to convert the carbon content to biomass, we multiply the former by a factor of $\sim 2$ \citep{MHT18} because the overwhelming majority of biomass on Earth occurs as land plants \citep{BPM18}. Therefore, the total biomass on Earth is estimated to be $M_\mathrm{bio,\oplus} \sim 1.1 \times 10^{15}$ kg.

We have plotted (\ref{RelBioM}) as a function of the brown dwarf mass in Figure \ref{FigBiomass} for the biomass densities delineated previously. It is seen that the maximum biomass that can be sustained is a monotonically decreasing function of $M_\mathrm{BD}$; this occurs because the radius and height of the habitable layer both decrease with the mass. It is also evident from Figure \ref{FigBiomass} that the maximum biomass is actually \emph{higher} than that of the Earth in all instances when the optimistic value for the biomass density ($\rho_\mathrm{max}$) is employed. However, if the lower limit for the biomass density ($\rho_\mathrm{min}$) is utilized, the resultant biomass is orders of magnitude smaller than Earth's biosphere, albeit still high viewed in absolute terms. The case with the mean biomass density ($\bar{\rho}$) is notable as straddles both regimes, i.e., $\delta_\mathrm{bio} > 1$ and $\delta_\mathrm{bio} < 1$, with the transition between them occurring at $M_\mathrm{BD} \approx 6.5 M_J$.

\subsection{Aerosols and the origin of life}
There is a great deal that remains unknown about the origin of life and the habitats in which it could have been actualized \citep{SMI08,Lu16,KM18}. Many of the common geochemical environments posited for the origin of life on Earth, such as hydrothermal vents, geothermal fields and beaches, are clearly unavailable in atmospheres. It has been hypothesized since the 1950s that aerosols represent viable sites for the initiation of abiogenesis \citep{Gold58}. This subject has witnessed some notable developments in recent times, as reviewed in \citet{DTTV} and \citet{GTV12}.

We will briefly highlight some of the salient advantages stemming from aerosols in a qualitative fashion, before tackling a couple of quantitative aspects later.
\begin{itemize}
    \item Observations and laboratory studies have revealed that aerosols with inverted micelle structures exist near water-air interfaces on Earth. Aerosols of this type comprise liquid water, minerals and small organic molecules enclosed within an organic film made up of fatty acids \citep{DTTV}.
    \item These structures are akin to vesicles, although the lipid bilayers that constitute the boundary in vesicles possess greater functionality. Vesicles have been posited to play an important role in the origin of protocells as they not only offer a natural compartmentalization scheme, but also permit the replication of biopolymers in their interiors \citep{CW10,BS14}.
    \item As the aerosols traverse the atmosphere, they should experience fluctuations in the ambient humidity. In brown dwarf atmospheres that may have patchy water clouds \citep{MMF14}, moving in and out of these clouds would induce a similar effect. The advantage of heterogeneity in relative humidity is that the onset of hydration-dehydration cycles is feasible \citep{Tuck02}. The importance of such wet-dry cycles is well established, as they enable the selection, concentration and polymerization of prebiotic monomers \citep{DD15,BSO18}.
    \item The synthesis of biopolymers is obviously an important step toward the origin of life. It has been shown that the formation of peptide bonds, which is disfavored on thermodynamic and kinetic grounds in aqueous environments, can take place at air-water interfaces such as those found in atmospheric aerosols \citep{GV12}. 
\end{itemize}
As noted above, aerosols that possess the inverted micelle structure might contribute to the origin of protocells. However, it is a well-known fact that cells divide. In order to mimic protocells in this regard, it is therefore necessary for the aerosols to be capable of fission. It was pointed out in \citet{DTV01} that the fission of pure aerosol particles is not possible on thermodynamic grounds because the free energy is already at a minimum, but splitting can occur in the case of aerosols with organic films. If the surface tension of the parent and two daughter aerosols is denoted by $\gamma_P$, $\gamma_1$ and $\gamma_2$, respectively, \citet{DTV01} found that fission could occur provided that
\begin{equation}
    \gamma_1 + \gamma_2 \zeta^2 < \gamma_P \left(1 + \zeta^3\right)^{2/3}, 
\end{equation}
where $\zeta = r_2/r_1$ is the ratio of the radii of the daughter aerosols and has a range of $\left(0,1\right)$. Therefore, symmetric division with $\zeta \approx 1$ requires $\gamma_1 + \gamma_2 < 1.59 \gamma_P$, which is somewhat unrealistic as it calls for sizable changes in the surface tension of daughter particles. However, asymmetric division with $0.1 < \zeta < 1$ is more feasible.

Lastly, we turn our attention to the issue of spatial and temporal scales for abiogenesis. Conventionally, abiogenesis has been envisioned as the successful outcome of a very large number of random ``trials''. Both intuitively and mathematically, it can be shown that the probability of abiogenesis is linearly proportional to the number of trials ($N_T$) conducted \citep{deDu05}. However, this ansatz is based on the premise that the \emph{mean quality} of each trial is comparable across different worlds. As we will henceforth compare trials in aerosol reactors in Earth's atmosphere with those in atmospheric habitable zones of brown dwarfs, we make the assumption that the mean quality per trial is similar in both instances. In assessing $N_T$, it must be noted that we have to account for both spatial and temporal factors. We begin by evaluating the spatial aspect. In an influential mathematical model, \citet{Dys99} proposed that $\sim 10^{10}$ droplets were necessary for the emergence of a low-entropy state from a disordered population of prebiotic monomers, analogous to the ferromagnetic phase transition.

\begin{figure}
\includegraphics[width=7.5cm]{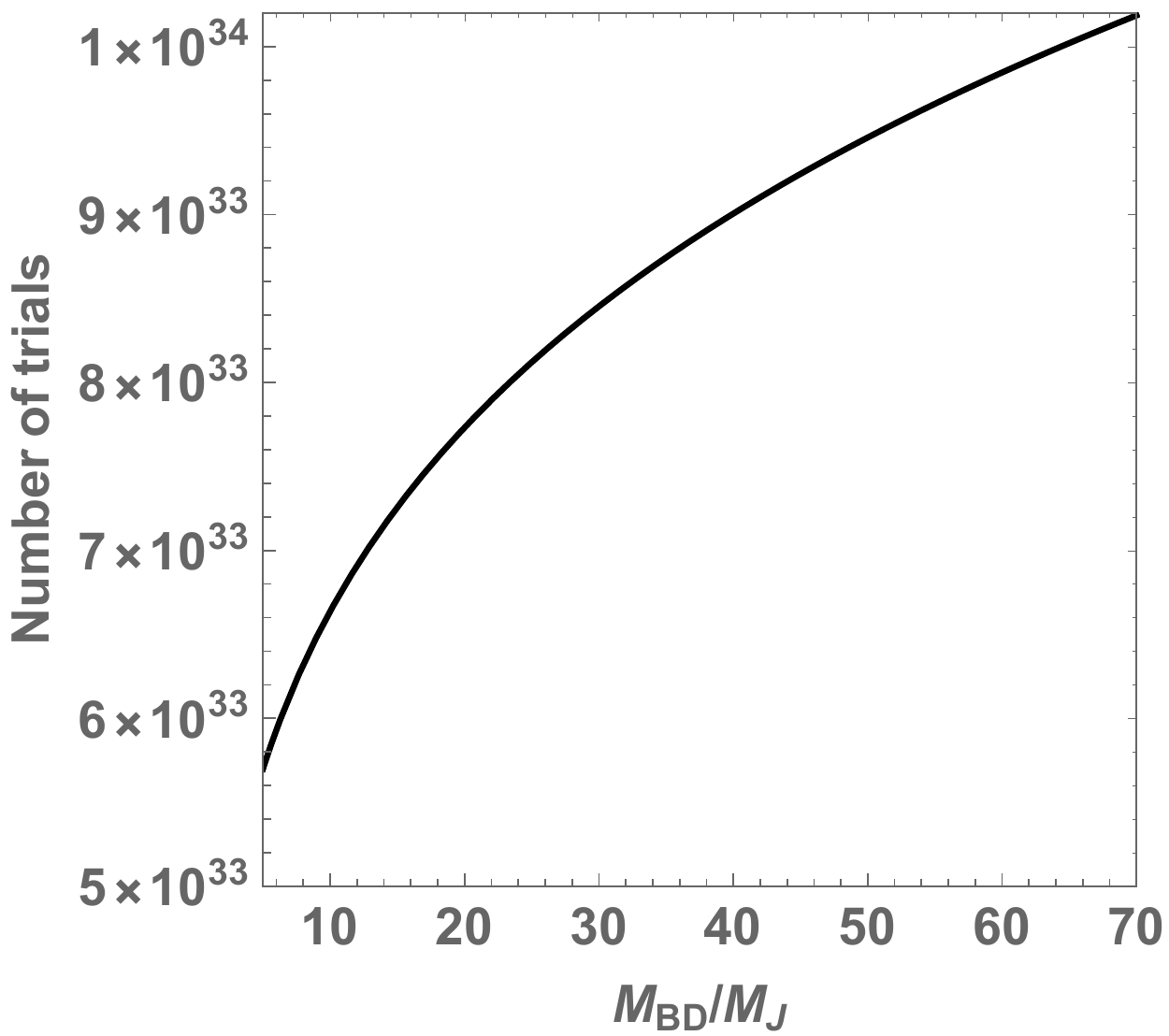} \\
\caption{The number of abiogenesis trials in aerosols that can occur in the atmospheric habitable zones of brown dwarfs as a function of the brown dwarf mass ($M_\mathrm{BD}$) expressed in units of Jupiter's mass ($M_J$). We have assumed the fiducial values of $n_a \sim 10^7$ m$^{-3}$ and $\tau_C \sim 100$ s in (\ref{NumT}). In comparison, the number of trials that took place in Earth's atmospheric aerosols at the time life originated is $\sim 10^{31}$.}
\label{FigTrials}
\end{figure}

The number density of aerosols ($n_a$) in cool brown dwarfs will clearly be spatially variable, but we adopt a fiducial value of $\sim 10^7$ m$^{-3}$ as it is compatible with observations of Jupiter's atmosphere (see Section \ref{SSecBMAHZ}). Therefore, the number of spatial trials possible in the atmospheric habitable zone is $V_\mathrm{BD} n_a/10^{10}$. Next, we turn to the temporal element. \citet{Dys99} suggested that the duration of the transition to an ordered state could be collectively viewed as constituting a single ``cycle''. Let us denote the time required for an individual cycle by $\tau_C$. Thus, the number of temporal trials over the entire habitability interval is given by $\tau_\mathrm{BD}/\tau_C$. Using the estimate for the number of temporal trials from \citet{Tuck02} in conjunction with the timescale for abiogenesis on Earth based on the current fossil and phylogenetic evidence \citep{Dodd17,BPC18}, we obtain a fiducial value of $\sim 100$ s for the Earth after diving the latter by the former. By combining the prior factors together, we obtain
\begin{eqnarray}\label{NumT}
N_T \sim 4 \times 10^{33}\,\left(\frac{M_\mathrm{BD}}{M_J}\right)^{0.22} \left(\frac{n_a}{10^7\,\mathrm{m^{-3}}}\right) \left(\frac{\tau_C}{10^2\,\mathrm{s}}\right)^{-1}.
\end{eqnarray}
In contrast, it has been estimated that $N_T \lesssim 10^{31}$ if one selects the time at which the first evidence for life on Earth appears in the geological record \citep{Tuck02}. Therefore, we find that the maximum number of trials available to brown dwarfs over their entire habitable period is possibly $\sim 3$ orders of magnitude higher with respect to Earth's atmosphere at the time of abiogenesis, as seen from Figure \ref{FigTrials} where $N_T$ is plotted as a function of $M_\mathrm{BD}$ for the choices $n_a \sim 10^7$ m$^{-3}$ and $\tau_C \sim 100$ s.

\subsection{Nutrient and energy availability}
Apart from the necessity of liquid water, both energy and nutrients are known to be essential preconditions for the origin and sustenance of biospheres. Modeling the energy and nutrient inventories is a complex endeavor, owing to which we shall focus on a few select aspects. 

\subsubsection{Bioessential elements}\label{SSSecBE}
The chief bioessential elements are CHNOPS. We will not explicitly deal with carbon, hydrogen and oxygen as they are clearly accessible in the form of atmospheric methane and water, respectively. The issue of sulfur metabolism has been investigated in the context of Venus \citep{SGA04,LMS18}. In particular, it has been suggested that the analogs of \emph{Acidithiobacillus ferrooxidans} could exist on Venus. In anaerobic conditions, \emph{A. ferrooxidans} uses Fe$^{3+}$ as an electron acceptor and oxidizes elemental sulfur to generate products such as sulfuric acid. 

With regards to sulfur, one of the key points worth bearing in mind is that virtually all of the sulfur inventory in brown dwarf atmospheres is anticipated to exist in the form of hydrogen sulfide \citep{VLF06}. In addition, sulfide clouds comprising ZnS, MnS and Na$_2$S are expected to exist. Therefore, as the most common version of sulfur exists in the form of sulfide, it would make sense for putative metabolic pathways to utilize the sulfide anion. One of the best known metabolic pathways involving H$_2$S as the reactant is anoxygenic photosynthesis, with \emph{Chlorobiaceae} (green sulfur bacteria) representing a classic example. A schematic representation of this pathway is
\begin{equation}
   \mathrm{CO_2} + 2\mathrm{H_2S} \rightarrow \mathrm{CH_2O} + \mathrm{H_2O} + 2\mathrm{S}.
\end{equation}
However, two difficulties encountered by this reaction are the availability of CO$_2$ and photons of suitable energy; note, however, that sufficient CO$_2$ may exist in the lower atmosphere \citep{BRH13} and we address photon availability in Section \ref{SSSecEM}.

Sulfur oxidizing bacteria represent another interesting candidate \citep{NS97}, but they typically require electron acceptors such as oxygen and nitrate (NO$_3^-$), neither of which are abundant in cool brown dwarf atmospheres. On the other hand, metal oxides present in the atmosphere might instead play the role of reactants and undergo reduction. Before moving on, we note that trace quantities of FeS could exist, but high abundances are unlikely as iron is expected to condense first to yield Fe clouds \citep{VLF06,VLF10}. Note that iron-sulfur compounds are highly important in life-as-we-know-it, because they constitute the basis of iron-sulfur clusters found in many key proteins \citep{BCJ08}, and may have placed a vital role in the origin of life \citep{Wac90}.

Next, we turn our attention to nitrogen. It is well-known that ammonia (NH$_3$) is a prominent component of the atmospheres of cool brown dwarfs. The existence of ammonia obviates the necessity for biological nitrogen fixation (biosynthesis of ammonia) by diazotrophs. The availability of ammonia, or ammonium (NH$_4^+$) in some instances, is also vital from the standpoint of Earth's nitrogen cycle because aerobic ammonia oxidizing bacteria convert ammonia to nitrite through a compound reaction involving hydroxylamine \citep{KS01} illustrated below:
\begin{eqnarray}
&& 2\mathrm{H^+} + \mathrm{NH_3} + 2\mathrm{e^-} + \mathrm{O_2} \rightarrow \mathrm{NH_2OH} + \mathrm{H_2O} \nonumber \\
&& \mathrm{NH_2OH} + \mathrm{H_2O} \rightarrow \mathrm{NO_2^-} + 5\mathrm{H^+} +  4\mathrm{e^-}.
\end{eqnarray}
Subsequently, nitrite is converted to nitrate by nitrite oxidizing bacteria; in turn, nitrate is acquired by biological organisms as it constitutes a vital nutrient. In the absence of oxygen, anammox (anaerobic ammonium oxidation) bacteria can oxidize ammonia to yield N$_2$. On account of the presence of atmospheric ammonia, microbial metabolic pathways along the above lines may be feasible, but they will necessitate the availability of suitable oxidants such as oxygen or nitrite. 

The last major bioessential element that needs to be addressed is phosphorus \citep{West87}. In the case of both modern and Proterozoic Earth, dissolved phosphorus (in the form of phosphates) is often regarded as the ultimate limiting nutrient over long timescales \citep{Tyrr99,SG06,LS18}. The same conclusion is expected to hold true for certain classes of exoplanets such as ocean worlds \citep{MaLi19}. One of the chief issues with phosphorus availability is that most phosphorus on Earth exists as mineral phosphates, which are characterized by very low solubilities; the best known example in this category is fluorapatite (Ca$_5$(PO4)$_3$F). At a pH of $7$ and a temperature of $307$ K, the solubility of fluorapatite in pure water is $\sim 3 \times 10^{-3}$ g/L \citep{McC68}.

We will now pivot to phosphorus content in cool brown dwarfs. At the effective temperatures considered herein, under the assumption of equilibrium chemistry, most of the phosphorus should exist in the form of tetraphosphorus hexaoxide (P$_4$O$_6$) as per theoretical calculations \citep{FL94}. Of more interest to us is the compound ammonium dihydrogen phosphate (NH$_4$H$_2$PO$_4$). Its condensation temperature ($T_c$) is estimated from equation (52) of \citet{VLF06} as:
\begin{equation}
    T_c \approx \frac{10^4\,\mathrm{K}}{30 - 0.2 \left(11 \log P_T + 15 [X/H]\right)},
\end{equation}
where $P_T$ is the pressure (in units of bar) and $[X/H]$ quantifies the metallicity. As an example, for a brown dwarf with solar metallicity at an altitude where the pressure is $\sim 1$ bar, the condensation temperature is found to be $T_c \sim 333$ K. It is therefore apparent that cool brown dwarfs might host clouds of NH$_4$H$_2$PO$_4$. Based on the observed L band spectra of the Y dwarf WISE J085510.83-071442.5, \citet{MSA18} suggested that the detected obscuration of near-infrared flux could be explained through the presence of NH$_4$H$_2$PO$_4$ clouds at a pressure of $\sim 10$ bar. 

As the effective temperature of WISE J085510.83-071442.5 ($\sim 250$ K) is within the thermal range considered herein, it appears likely that other brown dwarfs with this effective temperature also possess NH$_4$H$_2$PO$_4$. The reason we have underscored the existence of NH$_4$H$_2$PO$_4$ has to do with the issue of phosphorus limitation delineated above. The first, and perhaps the most important point, to appreciate is that NH$_4$H$_2$PO$_4$ is very soluble in water, thereby yielding the dihydrogen phosphate anion that may be utilized by putative organisms. At room temperature ($\sim 300$ K), the solubility of ammonium dihydrogen phosphate is $\sim 4 \times 10^2$ g/L \citep{Lide07}, about $10^5$ times higher than fluorapatite.

Second, we note that NH$_4$H$_2$PO$_4$ has been widely employed in laboratory experiments of prebiotic chemistry. A few salient examples are listed below.
\begin{itemize}
    \item In the 1960s and 1970s, several studies established that the phosphorylation of nucleosides to yield nucleotides and their oligomers (precursors of nucleic acids) was facilitated through the addition of ammonium dihydrogen phosphate and heating the mixtures \citep{PM65,LO71,HH74,OSS76}.
    \item Glycerol phosphates, which are important precursors of complex lipids that comprise cell membranes, can be synthesized by heating a mixture of glycerol and ammonium dihydrogen phosphate \citep{ENE79,DO80}. It should also be noted that the synthesis of phosphate amphiphiles (analogous to phospholipids in cell membranes) has been facilitated by using NH$_4$H$_2$PO$_4$ \citep{PS11}.
    \item In view of the ubiquity of polyphosphates (e.g., adenosine triphosphate) in biology, by drawing upon prior analyses, \citet{KM95} pointed out the fact that such compounds could (in)directly be generated by heating NH$_4$H$_2$PO$_4$ at fairly moderate temperatures of $\sim 333$-$373$ K.
\end{itemize}
Although a number of laboratory experiments employed NH$_4$H$_2$PO$_4$, they either did not endeavor or were unable to identify plausible sources for this compound \citep{KM95}. The latter aspect is not surprising because ammonium phosphates (including dihydrogen phosphate), despite their intrinsic advantages, were unlikely to have been available on Hadean Earth \citep{PGH17}. In contrast, as we have seen previously, cool brown dwarfs may have ammonium dihydrogen phosphate clouds at pressures of $\sim 1$-$10$ bar that constitute potentially viable sources of this compound.

Finally, we note that a number of other bioessential elements are predicted to exist in brown dwarf atmospheres such as manganese and iron. Hence, nutrient limitation vis-\`a-vis these elements might not necessarily pose a serious issue. However, we are not aware of any empirical or theoretical constraints on the abundance of molybdenum (Mo) or tungsten (W) in brown dwarf atmospheres.\footnote{The same issue also applies to the Venusian atmosphere, as pointed out by \citet{LL18}.} Molybdenum, in particular, is an essential component of many enzymes in prokaryotes and eukaryotes \citep{SMR09}, with the most notable being the nitrogenases that reduce nitrogen to yield ammonia; certain prokaryotes have evolved to use tungsten in lieu of molybdenum \citep{Hil02}. Hence, if Mo (or W) is a bioessential element insofar as life-as-we-know-it is concerned, it is important to gauge the abundance of this element in brown dwarf atmospheres. 

\subsubsection{Electromagnetic energy}\label{SSSecEM}
There are a number of energy sources that are accessible for prebiotic synthesis as well as putative biospheres in brown dwarf atmospheres.\footnote{In connection with prebiotic synthesis, owing to qualitative similarities between the atmospheres of cool brown dwarfs and Jupiter, it is conceivable that some prebiotic compounds listed in Section \ref{SecIntro} for the latter may also be synthesized in the former.} Some examples include cosmic rays, radioactivity, lightning and chemical energy \citep{DW10,LiLo19}. 

Instead of quantifying the fluxes for all these sources, we will focus only on electromagnetic radiation. On Earth, solar radiation constitutes the primary source of energy. It is therefore not surprising that photosynthesis is responsible for the majority of carbon fixation and biomass on Earth \citep{BPM18}. The emergence of photosynthesis was one of the major evolutionary transitions in our planet's history \citep{Knoll15}. On account of these reasons, we opt to analyze the availability of photosynthetically active radiation (PAR) in the atmospheres of cool brown dwarfs. 

Before proceeding further, a comment regarding the nature of photosynthesis is in order. The most productive version of photosynthesis on Earth is oxygenic photosynthesis, but its feasibility is probably lowered for primarily anoxic atmospheres. Instead, in the H$_2$-dominated atmospheres of brown dwarfs, it is plausible that ``hydrogenic photosynthesis'' is a viable pathway \citep{BSZ14}. The net reaction is expressible as
\begin{equation}\label{HydroPho}
    \mathrm{CH_4} + \mathrm{H_2O} \rightarrow \mathrm{CH_2O} + 2\mathrm{H_2},
\end{equation}
which is more transparent when separated into the two constituent half-reactions given by
\begin{eqnarray}
&& \mathrm{CH_4} + \mathrm{H_2O} \rightarrow \mathrm{CH_2O} + 4\mathrm{H^+} + 4\mathrm{e^-} \nonumber \\
&& \,\, 4\mathrm{H^+} + 4\mathrm{e^-} \rightarrow 2\mathrm{H_2}.
\end{eqnarray}
One of the major benefits underlying hydrogenic photosythesis is that both methane and water are abundant in cool brown dwarf atmospheres \citep{LF02,CKG11,ZM14}. There are a number of possible advantages stemming from hydrogenic photosynthesis that were reviewed in \citet{BSZ14}. Two of the most pertinent ones are: (a) the energy required for synthesizing a given quantity of biomass is $\sim 5$-$10$ times lower than oxygenic photosynthesis, and (b) the longest wavelength suitable for hydrogenic photosynthesis is $1.5$ $\mu$m, whereas conventional oxygenic photosynthesis ostensibly requires photons at wavelengths of $\lesssim 750$ nm \citep{NMS18}.

We begin our analysis by considering free-floating brown dwarfs that do not receive any radiation from external sources. As the atmospheric altitudes under consideration are smaller than $R_\mathrm{BD}$, to leading order we can model the maximum available PAR flux by the blackbody flux emitted in the appropriate wavelength range. The photon flux ($\Phi_\mathrm{BD}$) for the blackbody is
\begin{equation}\label{PhFluxBD}
\Phi_\mathrm{BD} = \int_{\lambda_\mathrm{min}}^{\lambda_\mathrm{max}} \frac{2c}{\lambda^4}\left[\exp\left(\frac{h c}{\lambda k_B T_\mathrm{eff}}\right)-1\right]^{-1}\,d\lambda,
\end{equation}
where we work with $T_\mathrm{eff} \approx 300$ K since we are analyzing brown dwarfs with effective temperatures in the range $\sim 250$-$350$ K. The maximum ($\lambda_\mathrm{max}$) and minimum ($\lambda_\mathrm{min}$) photon wavelengths corresponding to PAR are difficult to determine for other worlds, and will be addressed shortly hereafter. For now, we adopt the limits $\lambda_\mathrm{min} \approx 0.35$ $\mu$m and $\lambda_\mathrm{max} \approx 1.1$ $\mu$m as these wavelengths are compatible with those used by anoxygenic photoautotrophs on Earth \citep{KSG07,RLR18}. In our subsequent analysis, we will hold $\lambda_\mathrm{min}$ fixed as this is consistent with photosynthesis being inhibited by UV radiation \citep{Holl02,CBB07,CaG12}.

After substituting the above values into (\ref{PhFluxBD}), we end up with $\Phi_\mathrm{BD} \approx 1.2 \times 10^6$ m$^{-2}$ s$^{-1}$. In contrast, the minimum photon flux required for photosynthetic organisms on Earth is $\Phi_c \approx 1.2 \times 10^{16}$ m$^{-2}$ s$^{-1}$ \citep{RKB00,WoRa02}. This lower bound been explained through physicochemical constraints imposed by H$^+$ leakage, charge recombination, and protein turnover. Hence, as per the conventional PAR limits, it is impossible for photosynthesis to function in the atmospheres of free-floating brown dwarfs. 

Next, it is instructive to consider the total emitted photon flux ($\Phi_\mathrm{tot}$) given by
\begin{equation}
   \Phi_\mathrm{tot} \approx 4.8 c \left(\frac{k_B T_\mathrm{eff}}{h c}\right)^3. 
\end{equation}
Evaluating $\Phi_\mathrm{tot}$ for $T_\mathrm{eff} \approx 300$ K, we obtain $\Phi_\mathrm{tot} \approx 1.3 \times 10^{22}$ m$^{-2}$ s$^{-1}$. Interestingly, this value is $\sim 6.5$ times higher than the total solar photon flux incident on the Earth. Therefore, insofar as total photon flux is concerned, the lower atmospheres of brown dwarfs could receive a higher photon flux relative to the Earth. The chief difference, however, is that most of the photons are emitted at wavelengths of $\sim 10$ $\mu$m.

We will now examine the maximum wavelength that is necessary to facilitate \emph{oxygenic} photosynthesis. \citet{WoRa02} (see also \citealt{HR83}) suggested that oxygenic photosynthesis could function at longer wavelengths by harnessing a higher number of long-wavelength photons to carry out the fixation of CO$_2$, which is consequently accompanied by the generation of O$_2$. In the event that multiple ($> 2$) photons are required per electron, it is important to recognize that certain components of the hypothesized photosynthetic apparatus (e.g., water-oxidation complex) will probably differ from those found on Earth in several key respects, such as having a multiple Z-scheme \citep{KST07}. As the number of photons required increases, the evolution of a more intricate mechanism with its attendant catalytic macromolecules becomes necessary. It is impossible to ascertain whether the evolution of this apparatus is practically feasible; however, insofar as physics is concerned, there are no manifestly evident constraints that appear to rule it out. 

The following theoretical relationship has been proposed for multi-photon oxygenic photosynthesis \citep{WoRa02,KST07}:
\begin{equation}\label{chidef}
    \chi \approx 2\left(\frac{\lambda_\mathrm{max}}{0.7 \mathrm{\mu m}}\right),
\end{equation}
where $\chi$ denotes the number of photons required per electron transfer in oxygenic photosynthesis; the normalization is specified based on oxygenic photoautotrophs on Earth. In this event, the minimum photon flux required becomes $\left(\chi/2\right) \Phi_c$. By imposing the constraint that $\Phi_\mathrm{BD}$ should equal this value and invoking (\ref{PhFluxBD}), we obtain  $\lambda_\mathrm{max} \approx 2.7$ $\mu$m and $\chi \approx 7.8$. In other words, in order for oxygenic photosynthesis to occur, a unique photosystem based on eight photons per electron would presumably be necessary; for $\chi = 8$ (the given $\chi$ should be an integer), we find $\lambda_\mathrm{max} = 2.8$ $\mu$m.

At this stage, some crucial caveats should be mentioned. First, our analysis presupposed that all photons in the range $\lambda_\mathrm{min} < \lambda < \lambda_\mathrm{max}$ will reach the atmospheric layer where the biota are present. Second, the minimum photon flux represents an ideal limit as it assumes that $100\%$ absorption by the photosystem(s). Third, if eight photons are required per electron transfer, the absorption of an equal number of photons at higher energies could cause overheating and disrupt the photosynthetic apparatus. Finally, we note that thermodynamics itself places strict constraints on the efficiency at which the radiation can be utilized. Suppose that the ambient temperature in the atmospheric layer is $T_a$. The Carnot efficiency ($\eta_C$) serves as an upper bound in most (but not all) instances.\footnote{Note that the efficiency of a Carnot engine presented in (\ref{Carnot}) is not always correct \citep{CA75}.} The standard expression for $\eta_C$ is
\begin{equation}\label{Carnot}
    \eta_C = 1 - \frac{T_a}{T_\mathrm{eff}}.
\end{equation}
Considering $T_a \approx 280 K$ and $T_\mathrm{eff} \approx 310$ K, we see that the Carnot efficiency is only around $10\%$. In fact, the above formula predicts that extracting work becomes impossible when $T_a > T_\mathrm{eff}$. A more accurate treatment necessitates calculating the \emph{exergy} in the PAR range \citep{DB17,Sch19}, which goes beyond the scope of this paper. The Carnot efficiency constraint becomes relatively unimportant when it comes to cool brown dwarfs that are companions of stars.

On account of the above limitations, it is very plausible that the above values of $\lambda_\mathrm{max}$ and $\chi$ derived constitute lower bounds. On the other hand, we note that the photosynthetic machinery in other worlds might possess a higher efficiency and functionality with respect to Earth-based organisms as a result of having evolved in low-light conditions. In addition, our derivation ignored the possibility of PAR being derived from a host star (if one exists) or even high-energy astrophysical objects such as active galactic nuclei \citep{LGB19}.

It is worth examining the PAR accessible from the host star in more detail. By modelling the star as a blackbody, it is possible to estimate the critical orbital radius ($a_c$) at which the PAR flux becomes equal to $\Phi_c$ introduced previously. After simplifying the resultant expression, we obtain
\begin{equation}\label{CritDist}
 a_c \sim 316\,\mathrm{AU}\,\left(\frac{L_\star}{L_\odot}\right)^{1/2}\left(\frac{T_\star}{T_\odot}\right)^{-1/2} \sqrt{\frac{\mathcal{I}(T_\star)}{\mathcal{I}(T_\odot)}},   
\end{equation}
where $L_\star$ and $T_\star$ are the bolometric luminosity and temperature of the host star, whereas $\mathcal{I}$ is given by
\begin{equation}
\mathcal{I}(T_\star) \approx \int_{\ell_1(T_\star)}^{\ell_2(T_\star)} \frac{x'^2\,dx'}{\exp\left(x'\right) - 1},    
\end{equation}
where we have introduced $\ell_1(T_\star) \approx 3.32 \left(T_\star/T_\odot\right)^{-1}$ and $\ell_2(T_\star) \approx 7.12 \left(T_\star/T_\odot\right)^{-1}$. In deriving (\ref{CritDist}), we have opted for the conservative PAR range of $0.35$-$0.75$ $\mu$m. In actuality, as noted earlier, the maximal wavelength could increase if multi-photon schemes are viable. The only difference in this case is that $\mathcal{I}(T_\star)$ in (\ref{CritDist}) needs to be replaced by the function $\mathcal{K}(T_\star)$ defined as
\begin{equation}
\mathcal{K}(T_\star) \approx \frac{2}{\chi}\int_{2\ell_1(T_\star)/\chi}^{\ell_2(T_\star)} \frac{x'^2\,dx'}{\exp\left(x'\right) - 1},    
\end{equation}
where it should be recalled that $\chi$ denotes the number of photons used per electron transfer.

\begin{figure}
\includegraphics[width=7.5cm]{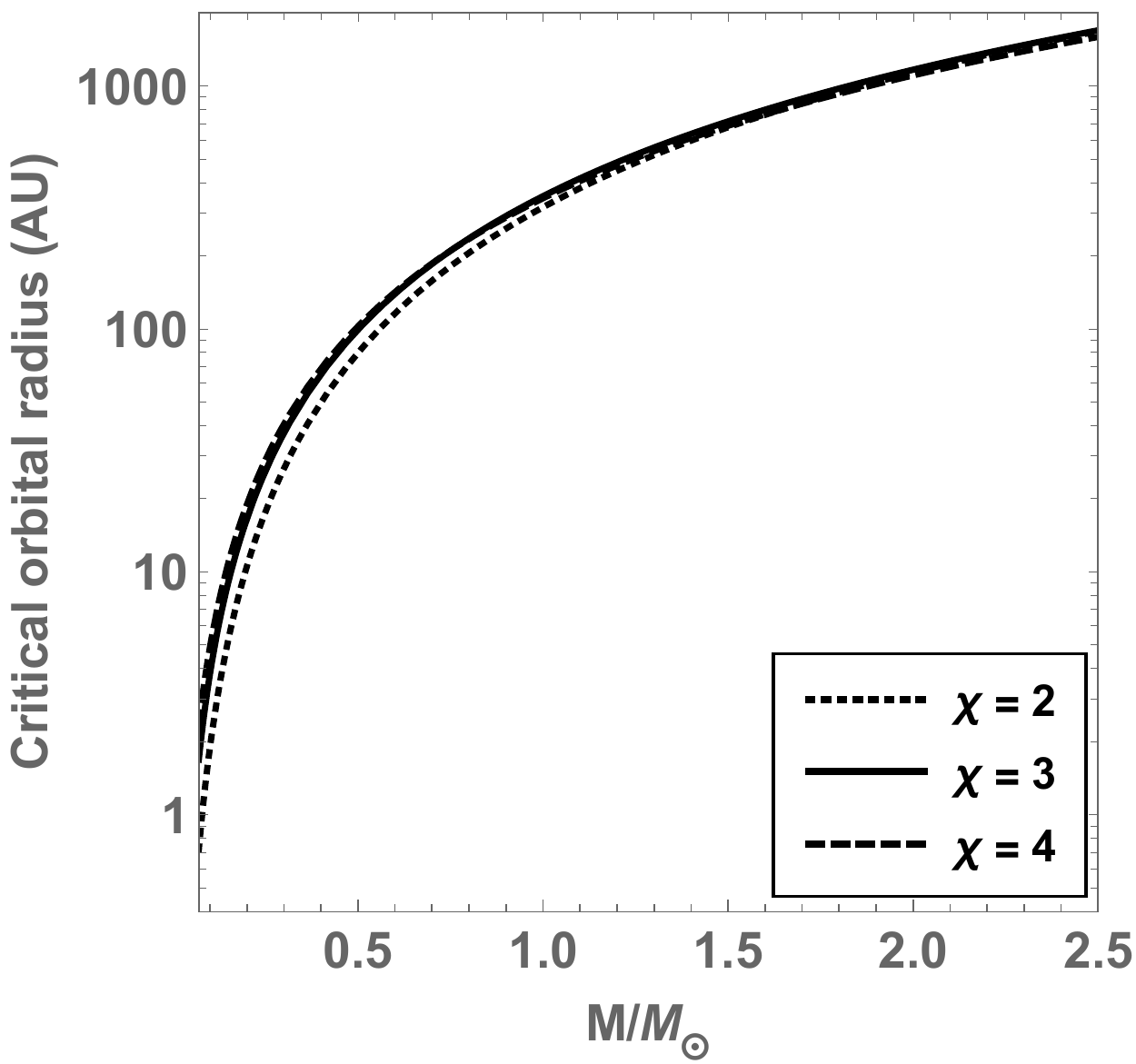} \\
\caption{The maximum orbital radius at which photosynthesis might be feasible as a function of the stellar mass ($M_\star$) in units of solar mass ($M_\odot$). The three curves correspond to the number of photons ($\chi$) utilized per electron transfer, where $\chi = 2$ for conventional oxygenic photosynthesis. }
\label{FigPhoto}
\end{figure}

In Figure \ref{FigPhoto}, we have plotted the critical orbital radius $a_c$ as a function of the stellar mass ($M_\star$) after employing empirical mass-radius and mass-temperature scalings described in \citet{Ling19} as well as the number of photons ($\chi$) involved in photosynthesis. From inspecting Figure \ref{FigPhoto}, we see that the number of photons utilized per electron transfer does not alter our results significantly for stars with $M_\star > M_\odot$. However, when we consider $M_\star \sim 0.1 M_\odot$, we find $a_c \approx 1.9$ AU for conventional photosynthesis ($\chi = 2$), whereas $a_c \approx 4.9$ AU for $\chi = 4$. Hence, the evolution of multi-photon schemes could enable an increase of the width of the photosynthesis ``zone'' around low-mass stars.\footnote{We have not explicitly investigated the role of stellar flares in powering photosynthesis, as their contribution is probably minimal for the majority of stars \citep{Ling19}.} 

Our estimate for $a_c$ in (\ref{CritDist}) is merely a loose upper bound because we have not taken the opacity of the brown dwarf atmosphere in the visible (or near-IR) into account. In reality, (\ref{CritDist}) must be multiplied by the factor $\exp\left(-\bar{\tau}_V\right)$, where $\bar{\tau}_V$ quantifies the optical depth of the brown dwarf atmosphere (until the atmospheric habitable zone is reached) in the appropriate wavelength range. Hence, it is possible that $a_c$ might be smaller by orders of magnitude but this is hard to determine \emph{a priori} since the optical depth is regulated by pressure and temperature, the abundances of various chemical species, and the wavelength \citep{MR15}. In the event that most of the PAR from the star is obscured by the upper atmosphere when it reaches the habitable zone, the outcome for photosynthesis is rendered analogous to free-floating brown dwarfs that we have previously analyzed in this Section.

\section{Detecting life in brown dwarf atmospheres}\label{SecLifeD}
We will briefly examine brown dwarf statistics, discuss potential biosignatures in the atmospheres of brown dwarfs, and the prospects for detecting them. 

\subsection{Brown dwarf statistics}\label{SSecBDSt}
Until this stage, we have not explicitly addressed the question of whether the brown dwarfs under consideration are situated around a host star, in binaries, or free-floating. We will briefly explore this issue, as it has implications for the search strategies. An inspection of (\ref{Teffrel}) reveals that achieving an effective temperature of $T_\mathrm{eff} \sim 300$ K within the current age of the Universe ($\sim 10^{10}$ yr) is feasible only for brown dwarfs with $M_\mathrm{BD} \lesssim 20 M_J$. Hence, broadly speaking, late-type T dwarfs and Y dwarfs are of primary interest to us. 

In the 2000s and 2010s, surveys of brown dwarf binaries have established the following properties: (a) their occurrence rate ($\lesssim 20\%$) is lower than the stellar population, (b) they are mostly found in tightly bound orbits at separations of a few AU, and (c) their mass ratio distribution has a sharp peak near unity and declines rapidly thereafter \citep{CSF03,BKC06,KH12}. In a recent study, \citet{FBB18} surveyed late-type T dwarfs and Y dwarfs and found that the binary fraction was $\sim 5.5\%$, with a peak in separation at $2.9$ AU and a power-law exponent of $6.1$ for the mass-ratio distribution. 

Next, we consider brown dwarf companions to stars. It has been suspected since the 1980s \citep{CWY88} that there is a paucity of brown dwarfs within a few AU of solar-type stars \citep{CWY88}, which came to be known as the brown dwarf ``desert'' \citep{MB00}. This was confirmed by a number of subsequent studies of FGKM stars \citep{EIK12,MG14,RMC16}, which found that the fraction of brown dwarfs at distances of $\sim 1$-$100$ AU was merely a few percent \citep{MH09,SSQ11,DHG12,CKI15}, whereas giant planets were relatively more abundant by a factor of $\lesssim 10$ \citep{GL06,LD07}. 

A statistical analysis of stars from various spectral classes at different ages concluded that the probability distribution function ($\mathcal{P}$) of substellar companions obeyed $\mathcal{P} \propto M^{-0.65} a^{-0.85}$, where $M$ and $a$ were the mass and orbital radius of the substellar object; it was found that the fraction of stars hosting $5$-$70\,M_J$ objects at $10$-$100$ AU was $\sim 1$-$3\%$ at $68\%$ confidence \citep{BET12}. Although close-in brown dwarfs are relatively scarce, it does not imply that they are completely absent. For instance, nine brown dwarf companions to solar-type stars have been identified at distances of $\sim 0.5$-$4$ AU based on data collected by the SOPHIE spectrograph \citep{WHS16}.

\subsection{Potential biosignatures}\label{SSecPotBS}
The biosignatures produced would be directly dependent on the putative organisms in question. For example, two of the most well-known biosignatures on Earth are molecular oxygen (O$_2$) and the ``red edge'' of vegetation, both of which are consequences of oxygenic photosynthesis \citep{SKP18}. 

First, we consider the presence of dead or decomposing organisms. In this scenario, in lieu of the organisms themselves their constituent biomolecules would play a prominent role. In the case of many Earth-based biomolecules, it is well-known that their peak absorption lies in the UV and visible regions. A review of many of these biomolecules can be found in \citet{LMS18}. Nucleic acids and proteins have peak absorbances at wavelengths of $260$ nm and $280$ nm, respectively. Iron-sulfur proteins, which play vital roles in redox reactions, are characterized by maximum absorption at wavelengths of $280$-$450$ nm. A number of biological pigments such as chlorophylls, carotenoids and pterins are strong absorbers at wavelengths $< 500$ nm. Thus, viewed collectively, we might expect to see an increase in the reflectance (i.e., decrease in absorption) spectra at wavelengths $\gtrsim 500$ nm. As this behavior has been documented for the Venusian atmosphere, it has led to suggestions that Venus' clouds might be harboring microbes \citep{LMS18}.

Next, let us turn our attention to live organisms. In principle, a number of microbial metabolisms are feasible as outlined in Section \ref{SSSecBE}, but some variant of photosynthesis comes across as a natural candidate due to its ubiquity and importance on Earth. Along the expected lines, the spectral red edge roughly coincides with $\lambda_\mathrm{max}$ on Earth. If we posit that a similar situation holds true for the photoautotrophic organisms in cool brown dwarf atmospheres, various possibilities open up depending on whether the brown dwarf is free-floating or bound as well as the nature of the photosynthetic pathway. 

We begin by tackling photosynthesis on free-floating brown dwarfs. As explained in Section \ref{SSSecEM}, $\lambda_\mathrm{max}$ for (oxygenic) photosynthesis is $\sim 2.8$ $\mu$m. If the minimum flux required for hydrogenic photosynthesis is comparable to its oxygenic counterpart, we can solve for $\lambda_\mathrm{max}$, with $0.7$ $\mu$m in the denominator of (\ref{chidef}) replaced by $1.5$ $\mu$m, after postulating that the biophysical process requires two photons of $1.5$ $\mu$m. Hence, we end up with $\lambda_\mathrm{max} \approx 2.6$ $\mu$m and $\chi \approx 3.5$; in other words, hydrogenic photosynthesis entailing $3$-$4$ photons might be feasible. For $\chi = 3$ and $\chi = 4$, the corresponding values of $\lambda_\mathrm{max}$ are approximately $2.25$ $\mu$m and $3.0$ $\mu$m, respectively. Therefore, for free-floating brown dwarfs, the manifestation of a ``red edge'' close to the outer boundary of the near-infrared (near-IR), i.e., at wavelengths of $\sim 2.3$-$3.0$ $\mu$m, comes across as being plausible.

Next, we turn our attention to brown dwarf companions around stars. As long as the criterion $a < a_c$ is satisfied, enough photons for photosynthesis (either hydrogenic, anoxygenic or oxygenic) should be accessible to photoautotrophs. Hence, if the exact analog of oxygenic photosynthesis exists, we would expect to see a spectral edge at $\sim 700$ nm. Instead, it is more plausible that the spectral edge will be manifested in the near-IR at wavelengths of $\lesssim 1.5$ $\mu$m, as both anoxygenic and hydrogenic photoautotrophs can utilize such wavelengths \citep{KST07,BSZ14}.

Apart from the photosynthetic spectral edge, the detection of biosignature gases produced by oxygenic photosynthesis is challenging. The oxygen thus produced would react quickly with reduced gases (that are abundant) unless it is generated in large quantities. Ammonia is a potential byproduct of hydrogenic photosynthesis \citep{BSZ14}, but it will be very challenging to distinguish between biotic and abiotic NH$_3$, given that the latter is plentiful in cool brown dwarfs. However, an interesting avenue for possibly identifying hydrogenic photosynthesis stems from noting that methane is depleted in accordance with (\ref{HydroPho}). 

Hence, if there is a mismatch between the abundance of methane inferred through observations and that determined by theory using only abiotic sources and sinks, it might be indicative of biological activity. Cooler atmospheres exhibit stronger signs of disequilibrium and, in principle, the diagnosis of chemical disequilibrium can be undertaken via the analysis of second eclipse spectra \citep{LY13,KTBC}; see also \citet{KTG18}. The spectra of the Y dwarf WISE J085510.83-071442.5 are compatible with an under-abundance of methane \citep{MSA18}, but we emphasize that this discrepancy (if it exists) is explainable via abiotic mechanisms. 

Needless to say, even the detection of such features is not indicative of life because it may instead arise from false positives. For instance, several abiotic materials such as dust, salts and polymers have been argued to explain the reduction in Venusian albedo observed at $\lesssim 500$ nm, with sulfur dioxide and iron chloride being two notable candidates \citep{ZKM81}. Likewise, minerals such as cinnabar (HgS) display sharp spectral edges that are reminiscent of the red edge of vegetation, albeit not at the same location \citep{STSF}. Along similar lines, it is conceivable that some of the spectral edges elucidated above overlap with those produced by abiotic substances. 

Second, as the atmospheres of brown dwarfs comprise layers of clouds, their existence will hinder measurements. However, in the event that the cloud cover is patchy - a feature that has been confirmed for some brown dwarfs \citep{RJL12} - time-resolved spectra might permit the identification of spectral features. Microscopic organisms may also experience horizontal and vertical transport due to atmospheric circulation \citep{SK13,ZS14,AKM17}, and could therefore be transported to regions with lower opacity, thereby presumably rendering their spectral features more discernible.

\subsection{Detectability of biosignatures}
As remarked previously, brown dwarfs can either exist on their own (i.e., free-floating) or as stellar companions. Observing free-floating brown dwarfs is advantageous from the standpoint of not having to concern ourselves with resolving their spectra by subtracting the contribution from the host star. 

On the other hand, the emission peak of brown dwarfs at $T_\mathrm{eff} \sim 300$ K is at $\sim 10$ $\mu$m, which typically entails observations undertaken in the mid-IR. For example, detailed spectra of WISE J085510.83−071442.5 (with $T_\mathrm{eff} \sim 250$ K) have been obtained in the L and M bands, corresponding to wavelength ranges of $3.4$-$4.14$ $\mu$m and $4.5$-$5.1$ $\mu$m, respectively. Yet, as outlined in Section \ref{SSecPotBS}, several interesting biosignatures are expected to manifest at visible and near-IR wavelengths, where the brown dwarf is many orders of magnitude fainter. 

For cool Y dwarfs at distances of $\lesssim 20$ pc, the first upper limits on the abundances of gases such as CH$_4$, H$_2$O, NH$_3$, H$_2$S and CO$_2$ were obtained recently using the Wide Field Camera 3 instrument on the Hubble Space Telescope \citep{ZLSP19}. The Near InfraRed Spectrograph (NIRSpec) on the upcoming James Webb Space Telescope (JWST) operates over a wavelength range of $0.6$-$5$ $\mu$m.\footnote{\url{https://jwst.nasa.gov/nirspec.html}} Numerical simulations undertaken by \citet{ZLSP19} suggest that a signal-to-noise ratio (SNR) of $\sim 200$ in the J-band is achievable with only $\sim 15$ minutes of integration time for objects at distances of $\sim 10$ pc, implying that it represents a powerful tool for characterizing the atmospheres of Y dwarfs. Hence, searching for atmospheric biosignatures of free-floating cool brown dwarfs is viable with JWST. A quantitative estimate of the yield from JWST is described toward the end of the section.

Next, directing our attention to brown dwarf companions, it was noted in Section \ref{SSecBDSt} that there exists a brown dwarf desert relative to giant planets. At this stage, we recall that cool brown dwarfs and giant planets share several similarities, although there also exist appreciable differences (e.g., surface gravity). Hence, our subsequent discussion is also applicable to giant planets with masses $M > M_J$ that may possess aerial biospheres. We will adopt the scaling relations reviewed in \citet{Winn10} and \citet{FAD18} henceforth. 

Even though (\ref{RaBD}) is more accurate for brown dwarfs, we will utilize it to calculate the radius $R$ of both brown dwarfs and giant planets a few times more massive than Jupiter; we refer to them collectively as substellar objects. First, we note that the transit depth scales as $R^2$, implying that the transit depth of the substellar object is $\sim 10^2$ times higher with respect to an Earth-sized planet. Next, for transmission spectroscopy, the SNR manifests the scaling:
\begin{equation}
    \mathrm{SNR} \propto R \mathcal{H} \Delta t^{1/2},
\end{equation}
where $\Delta t$ is the integration time and $\mathcal{H}$ is the scale height of the substellar object. We have opted to hold the stellar properties constant as well as the instrument specifications. Now, if we wish to determine the integration time for a fixed SNR, we see that
\begin{equation}
    \Delta t \propto \left(\frac{a}{R \mathcal{H}}\right)^2.
\end{equation}
Using the fact that the scale height is $\mathcal{H} \sim k_B T_a/(\bar{m} g)$, the above expression reduces to
\begin{equation}
    \frac{\Delta t}{\Delta t_\oplus} \sim 2.85 \times 10^{-7}\,\left(\frac{M}{M_J}\right)^4 \left(\frac{T_a}{250\,\mathrm{K}}\right)^{-2},
\end{equation}
after using equation (2.51) of \citet{BL93} for $g$. Note that $\Delta t_\oplus$ denotes the integration time required for an Earth-like planet. Hence, if we choose $T_a \sim 250$ K and $M \sim 10\,M_J$, we see that the integration time required to achieve a particular SNR drops by three orders of magnitude for this substellar object. 

Next, if one wishes to study thermal emission (i.e., emission spectrum) from the substellar object, the SNR will scale as
\begin{equation}\label{SNRTE}
    \mathrm{SNR} \propto R^2 \Delta t^{1/2},
\end{equation}
where the other parameters are held constant. Now, as before, if we consider a fixed value of SNR, we end up with $\Delta t \propto R^{-4}$, which simplifies to
\begin{equation}\label{IntTE}
 \frac{\Delta t}{\Delta t_\oplus} \sim  6.3 \times 10^{-7}\,\left(\frac{M}{M_J}\right)^{4/3}.  
\end{equation}
Therefore, upon selecting $M \sim 10 M_J$, from the above formula we see that the integration time relative to an Earth-sized planet decreases by five orders of magnitude. 

Lastly, let us suppose that we are interested in direct imaging of a cool substellar object via reflected light. The contrast ratio scales as $R^2$, implying that it increases by a factor of $\sim 10^2$ in comparison to an Earth-like planet. The SNR ought to exhibit the same scaling as (\ref{SNRTE}) apart from an extra factor of $a^{-2}$, implying that the desired integration time is given by (\ref{IntTE}) when $a$ is held fixed. Hence, the corresponding integration time would be lowered by a factor of $\sim 10^5$ with respect to an Earth-sized planet, when we consider a substellar object with $M \sim 10 M_J$.

Thus, the basic conclusion to be drawn herein is that the integration times required are considerably reduced, implying that achieving a high SNR is orders of magnitude more feasible when dealing with substellar objects with masses $M \gtrsim 10 M_J$ relative to characterizing Earth-like exoplanets. We will now quantify the yield of cool brown dwarf companions to stars, whose atmospheres are analyzable by JWST.

Out to a distance of $d_\star \sim 100$ pc from Earth, there are $\sim 5 \times 10^5$ stars, of which $\sim 50\%$ are M-dwarfs with $M_\star \lesssim 0.2 M_\odot$ \citep{KWP13}. Considering an orbital radius of $\sim 5$-$30$ AU (with a geometric mean of $\sim 12$ AU), it can be assumed that the fraction of stars with brown dwarfs is on order of $1\%$ \citep{EIK12,DHG12}. If we further specialize to objects with $M < 20 M_J$, the yield must be further lowered by a factor of $\sim 3$ based on the substellar IMF specified in \citet{KMS19}. Thus, by combining all these factors, we find that $\sim 10^3$ cool brown dwarfs might be suitable for biosignature characterization by JWST.

We begin with the case of characterizing substellar objects via transmission spectroscopy. We make use of equation (4) of \citet{FAD18} derived for JWST and work with $\mathcal{H} \sim 20$ km \citep{SK13},\footnote{It must be noted that the scale height is not constant for brown dwarfs because it is dependent upon the pressure and composition \citep{MR15}.} $R \sim R_J$ ($R_J$ is Jupiter's radius) and $d_\star \sim 100$ pc. The resultant SNR for JWST is found to be
\begin{equation}
    \mathrm{SNR} \sim 3.3 \left(\frac{\Delta t}{1\,\mathrm{hr}}\right)^{1/2},
\end{equation}
implying that a moderately high SNR is achievable for cool substellar objects even with hours of integration time. Next, we turn our attention to detecting the emission spectrum from these objects. We make use of equation (7) of \citet{FAD18} for the above choice of parameters, and obtain
\begin{equation}
    \mathrm{SNR} \sim 2.5 \varepsilon \left(\frac{\Delta t}{1\,\mathrm{hr}}\right)^{1/2},
\end{equation}
where $\varepsilon$ embodies the relative depth of spectral features. Hence, if we choose an integration time of a few hours, a reasonable SNR is attainable. A potential issue with detecting emission from brown dwarf companions is that the starlight needs to be separated from the brown dwarf, owing to which it is probably easier to study free-floating brown dwarfs via this avenue.

The last mode of observation entails reflected light from substellar objects. However, owing to the $1/a^{2}$ dependence and the relative paucity of brown dwarfs at distances of a few AU, this method is disfavored compared to the previous two methods described herein. For $a \sim 10$ AU and Bond albedo of $\sim 0.2$, the required contrast ratio to differentiate the substellar object from the star is $\sim 10^{-10}$, which is challenging with state-of-the-art coronographs and starshades \citep{FAD18}.

\section{Conclusion}\label{SecConc}
The search for extraterrestrial life outside our Solar system is expected to play a major role in the near-future. Currently, virtually all theoretical and observational studies are geared toward finding atmospheric biosignatures of rocky planets in the habitable zones of their host stars. However, despite a few studies in the context of our Solar system, the potential for life in \emph{atmospheric} habitable zones (i.e., aerial biospheres) has mostly gone unappreciated, with perhaps the only noteworthy exception being \citet{YPB17}.

In this paper, we have therefore investigated the atmospheric habitability of cool brown dwarfs, as well as sub-brown dwarfs and giant planets, at an effective temperature of $\sim 250$-$300$ K. In Section \ref{SecMaxHZ}, we began by estimating that the maximum habitable volume encompassed by cool brown dwarf atmospheres is conceivably two orders of magnitude higher than the volume associated with Earth-like planets in the habitable zones of their host stars. The reasons for the higher habitable volume are the greater spatial volume and temporal duration that collectively offset the fact that stars are more numerous than brown dwarfs. 

As there are many facets of putative extraterrestrial aerial biospheres that have not been investigated hitherto, we explored some of the key aspects in Section \ref{SecLifeBD}. By drawing upon data for Earth's current aerial biosphere in conjunction with empirical constraints on other Solar system objects, we found that the biomass encapsulated in the atmospheric habitable zones of cool brown dwarfs might surpass the Earth's biomass under optimal conditions. Next, we highlighted the significance of aerosols as potential prebiotic reactors in facilitating the origin of life, and thereby showed that the number of abiogenesis ``trials'' in brown dwarf aerosols possibly exceeds that of the Earth at the time of life's appearance around $4$ Ga by a factor of $\sim 100$-$1000$.\footnote{Our hypothesis concerning the origination of life within the atmospheres of cool substellar atmospheres ignores the possibility of these worlds being seeded by way of interstellar panspermia, despite the fact that recent calculations suggest that it might be feasible \citep{BMMS,Li16,GLL18}.} 

We surveyed the bioessential elements accessible to putative organisms in the atmosphere. We directed most of our attention toward phosphorus, as it constitutes the limiting nutrient on Earth. We highlighted the formation of ammonium dihydrogen phosphate and how it could serve as a ready source of soluble phosphorus as well as yield a number of vital prebiotic compounds. Subsequently, we explored the prospects for photosynthesis on free-floating brown dwarfs. Despite the general paucity of photons, we hypothesized that photosynthesis could function via a multi-photon scheme with a maximum wavelength of $\sim 2.3$-$3.0$ $\mu$m, i.e., close to the near-IR outer boundary. In contrast, for brown dwarfs that are stellar companions, photon availability is not expected to be a major limiting factor in most instances and the spectral edge would occur at either visible or near-IR wavelengths depending on the stellar spectrum.

In Section \ref{SecLifeD}, we presented a brief survey of brown dwarf statistics and assessed the spectral biosignatures that can result from the presence of life. We proposed that the analog of the red edge of vegetation might occur, albeit at wavelengths in the near-IR; the exact location of the spectral edge is dependent on the metabolic pathway. Another possibility is that chemical disequilibrium could result from the depletion or generation of certain gases (e.g., methane) that is potentially detectable. In the case of cool substellar objects around stars, we demonstrated that the required integration time to achieve a high SNR is orders of magnitude smaller with respect to Earth-sized planets at roughly the same distance and effective temperature. We find that $\sim 10^3$ cool brown dwarfs may be investigated for biosignatures by JWST at distances $\lesssim 100$ pc, with an integration time of $\mathcal{O}(1)$ hr yielding a SNR of $\sim 5$.

Thus, viewed collectively, there is arguably a strong case to be made for seeking atmospheric biosignatures in cool brown dwarfs and sub-brown dwarfs.\footnote{If life is detected on these worlds someday, perhaps they will merit the sobriquet ``green dwarfs''. The word ``green'' is particularly apropos if the existence of chlorophyll-type photosynthetic pigments is revealed, as the green color of vegetation on Earth is a direct consequence of chlorophylls.} A major advantage with pursuing this line of enquiry is that even the non-detection of life will still provide us with an in-depth understanding of planetary atmospheres because such objects exhibit stellar composition but are otherwise akin to giant planets in their atmospheric physics and chemistry. Apart from observational surveys, laboratory experiments are needed to properly gauge whether life could exist in conditions mimicking these cool atmospheres and what types of biosignatures would be most prominent. Finally, laboratory experiments and observations must be supplanted with theoretical and numerical models that assist in making testable predictions and interpreting empirical results.

\acknowledgments
We thank the reviewer for the very comprehensive and insightful report that helped improve the quality of the paper. We also thank Freeman Dyson and Corey Powell for valuable comments regarding the paper. This work was supported in part by the Breakthrough Prize Foundation, Harvard University's Faculty of Arts and Sciences, and the Institute for Theory and Computation (ITC) at Harvard University.

%\bibliographystyle{aasjournal}
%\bibliography{BDLife}

\begin{thebibliography}{}
\expandafter\ifx\csname natexlab\endcsname\relax\def\natexlab#1{#1}\fi
\providecommand{\url}[1]{\href{#1}{#1}}

\bibitem[{{Amato} {et~al.}(2007){Amato}, {Parazols}, {Sancelme}, {Mailhot},
  {Laj}, \& {Delort}}]{APS07}
{Amato}, P., {Parazols}, M., {Sancelme}, M., {et~al.} 2007, Atmospheric
  Environ., 41, 8253

\bibitem[{{Apai} {et~al.}(2017){Apai}, {Karalidi}, {Marley}, {Yang}, {Flateau},
  {Metchev}, {Cowan}, {Buenzli}, {Burgasser}, {Radigan}, {Artigau}, \&
  {Lowrance}}]{AKM17}
{Apai}, D., {Karalidi}, T., {Marley}, M.~S., {et~al.} 2017, Science, 357, 683

\bibitem[{{Arney} {et~al.}(2016){Arney}, {Domagal-Goldman}, {Meadows}, {Wolf},
  {Schwieterman}, {Charnay}, {Claire}, {H{\'e}brard}, \& {Trainer}}]{ADG16}
{Arney}, G., {Domagal-Goldman}, S.~D., {Meadows}, V.~S., {et~al.} 2016,
  Astrobiology, 16, 873

\bibitem[{{Bailey}(2014)}]{Bai14}
{Bailey}, J. 2014, Publ. Astron. Soc. Aust., 31, e043

\bibitem[{{Bains} {et~al.}(2014){Bains}, {Seager}, \& {Zsom}}]{BSZ14}
{Bains}, W., {Seager}, S., \& {Zsom}, A. 2014, Life, 4, 716

\bibitem[{{Bains} {et~al.}(2015){Bains}, {Xiao}, \& {Yu}}]{BXY15}
{Bains}, W., {Xiao}, Y., \& {Yu}, C. 2015, Life, 5, 1054

\bibitem[{{Bandyopadhyay} {et~al.}(2008){Bandyopadhyay}, {Chandramouli}, \&
  {Johnson}}]{BCJ08}
{Bandyopadhyay}, S., {Chandramouli}, K., \& {Johnson}, M.~K. 2008, Biochem.
  Soc. Trans, 36, 1112

\bibitem[{{Bar-On} {et~al.}(2018){Bar-On}, {Phillips}, \& {Milo}}]{BPM18}
{Bar-On}, Y.~M., {Phillips}, R., \& {Milo}, R. 2018, Proc. Natl. Acad. Sci.
  USA, 115, 6506

\bibitem[{{Becker} {et~al.}(2018){Becker}, {Schneider}, {Okamura}, {Crisp},
  {Amatov}, {Dejmek}, \& {Carell}}]{BSO18}
{Becker}, S., {Schneider}, C., {Okamura}, H., {et~al.} 2018, Nat. Commun., 9,
  163

\bibitem[{{Belbruno} {et~al.}(2012){Belbruno}, {Moro-Mart{\'{\i}}n},
  {Malhotra}, \& {Savransky}}]{BMMS}
{Belbruno}, E., {Moro-Mart{\'{\i}}n}, A., {Malhotra}, R., \& {Savransky}, D.
  2012, Astrobiology, 12, 754

\bibitem[{{Betts} {et~al.}(2018){Betts}, {Puttick}, {Clark}, {Williams},
  {Donoghue}, \& {Pisani}}]{BPC18}
{Betts}, H.~C., {Puttick}, M.~N., {Clark}, J.~W., {et~al.} 2018, Nat. Ecol.
  Evol., 2, 1556

\bibitem[{{Bilger} {et~al.}(2013){Bilger}, {Rimmer}, \& {Helling}}]{BRH13}
{Bilger}, C., {Rimmer}, P., \& {Helling}, C. 2013, Mon. Not. R. Astron. Soc.,
  435, 1888

\bibitem[{{Blain} \& {Szostak}(2014)}]{BS14}
{Blain}, J.~C., \& {Szostak}, J.~W. 2014, Annu. Rev. Biochem., 83, 615

\bibitem[{{Bolmont} {et~al.}(2017){Bolmont}, {Selsis}, {Owen}, {Ribas},
  {Raymond}, {Leconte}, \& {Gillon}}]{BSO17}
{Bolmont}, E., {Selsis}, F., {Owen}, J.~E., {et~al.} 2017, Mon. Not. R. Astron.
  Soc., 464, 3728

\bibitem[{{Bowers} {et~al.}(2012){Bowers}, {McCubbin}, {Hallar}, \&
  {Fierer}}]{BMH12}
{Bowers}, R.~M., {McCubbin}, I.~B., {Hallar}, A.~G., \& {Fierer}, N. 2012,
  Atmospheric Environ., 50, 41

\bibitem[{{Brandt} {et~al.}(2014){Brandt}, {McElwain}, {Turner}, {Mede},
  {Spiegel}, {Kuzuhara}, {Schlieder}, {Wisniewski}, {Abe}, {Biller},
  {Brandner}, {Carson}, {Currie}, {Egner}, {Feldt}, {Golota}, {Goto}, {Grady},
  {Guyon}, {Hashimoto}, {Hayano}, {Hayashi}, {Hayashi}, {Henning}, {Hodapp},
  {Inutsuka}, {Ishii}, {Iye}, {Janson}, {Kandori}, {Knapp}, {Kudo}, {Kusakabe},
  {Kwon}, {Matsuo}, {Miyama}, {Morino}, {Moro-Mart{\'{\i}}n}, {Nishimura},
  {Pyo}, {Serabyn}, {Suto}, {Suzuki}, {Takami}, {Takato}, {Terada}, {Thalmann},
  {Tomono}, {Watanabe}, {Yamada}, {Takami}, {Usuda}, \& {Tamura}}]{BET12}
{Brandt}, T.~D., {McElwain}, M.~W., {Turner}, E.~L., {et~al.} 2014, Astrophys.
  J., 794, 159

\bibitem[{{Burgasser} {et~al.}(2006){Burgasser}, {Kirkpatrick}, {Cruz}, {Reid},
  {Leggett}, {Liebert}, {Burrows}, \& {Brown}}]{BKC06}
{Burgasser}, A.~J., {Kirkpatrick}, J.~D., {Cruz}, K.~L., {et~al.} 2006,
  Astrophys. J., Suppl. Ser., 166, 585

\bibitem[{{Burrows} {et~al.}(2001){Burrows}, {Hubbard}, {Lunine}, \&
  {Liebert}}]{BHL01}
{Burrows}, A., {Hubbard}, W.~B., {Lunine}, J.~I., \& {Liebert}, J. 2001, Rev.
  Mod. Phys., 73, 719

\bibitem[{{Burrows} \& {Liebert}(1993)}]{BL93}
{Burrows}, A., \& {Liebert}, J. 1993, Rev. Mod. Phys., 65, 301

\bibitem[{{Caballero}(2018)}]{Cab18}
{Caballero}, J.~A. 2018, Geosciences, 8, 362

\bibitem[{{Cable} {et~al.}(2012){Cable}, {H{\"o}rst}, {Hodyss}, {Beauchamp},
  {Smith}, \& {Willis}}]{CHH12}
{Cable}, M.~L., {H{\"o}rst}, S.~M., {Hodyss}, R., {et~al.} 2012, Chem. Rev.,
  112, 1882

\bibitem[{{Caldeira} \& {Kasting}(1992)}]{CK92}
{Caldeira}, K., \& {Kasting}, J.~F. 1992, Nature, 360, 721

\bibitem[{{Caldwell} {et~al.}(2007){Caldwell}, {Bornman}, {Ballar{\'e}},
  {Flint}, \& {Kulandaivelu}}]{CBB07}
{Caldwell}, M.~M., {Bornman}, J.~F., {Ballar{\'e}}, C.~L., {Flint}, S.~D., \&
  {Kulandaivelu}, G. 2007, Photochem. Photobiol. Sci., 6, 252

\bibitem[{{Campbell} {et~al.}(1988){Campbell}, {Walker}, \& {Yang}}]{CWY88}
{Campbell}, B., {Walker}, G.~A.~H., \& {Yang}, S. 1988, Astrophys. J., 331, 902

\bibitem[{{Castenholz} \& {Garcia-Pichel}(2012)}]{CaG12}
{Castenholz}, R.~W., \& {Garcia-Pichel}, F. 2012, in {Ecology of Cyanobacteria
  II}, ed. B.~A. {Whitton} (Springer), 481--499

\bibitem[{{Cheetham} {et~al.}(2015){Cheetham}, {Kraus}, {Ireland}, {Cieza},
  {Rizzuto}, \& {Tuthill}}]{CKI15}
{Cheetham}, A.~C., {Kraus}, A.~L., {Ireland}, M.~J., {et~al.} 2015, Astrophys.
  J., 813, 83

\bibitem[{{Chen} \& {Walde}(2010)}]{CW10}
{Chen}, I.~A., \& {Walde}, P. 2010, Cold Spring Harb. Perspect. Biol., 2,
  a002170

\bibitem[{{Chyba}(2000)}]{Chy00}
{Chyba}, C.~F. 2000, Nature, 403, 381

\bibitem[{{Clarke}(2014)}]{Cla14}
{Clarke}, A. 2014, Int. J. Astrobiol., 13, 141

\bibitem[{{Close} {et~al.}(2003){Close}, {Siegler}, {Freed}, \&
  {Biller}}]{CSF03}
{Close}, L.~M., {Siegler}, N., {Freed}, M., \& {Biller}, B. 2003, Astrophys.
  J., 587, 407

\bibitem[{{Cockell}(1999)}]{Cock99}
{Cockell}, C.~S. 1999, Planet. Space Sci., 47, 1487

\bibitem[{{Curzon} \& {Ahlborn}(1975)}]{CA75}
{Curzon}, F.~L., \& {Ahlborn}, B. 1975, Am. J. Phys., 43, 22

\bibitem[{{Cushing} {et~al.}(2011){Cushing}, {Kirkpatrick}, {Gelino},
  {Griffith}, {Skrutskie}, {Mainzer}, {Marsh}, {Beichman}, {Burgasser},
  {Prato}, {Simcoe}, {Marley}, {Saumon}, {Freedman}, {Eisenhardt}, \&
  {Wright}}]{CKG11}
{Cushing}, M.~C., {Kirkpatrick}, J.~D., {Gelino}, C.~R., {et~al.} 2011,
  Astrophys. J., 743, 50

\bibitem[{{Damer} \& {Deamer}(2015)}]{DD15}
{Damer}, B., \& {Deamer}, D. 2015, Life, 5, 872

\bibitem[{{Dartnell} {et~al.}(2015){Dartnell}, {Nordheim}, {Patel}, {Mason},
  {Coates}, \& {Jones}}]{DNP}
{Dartnell}, L.~R., {Nordheim}, T.~A., {Patel}, M.~R., {et~al.} 2015, Icarus,
  257, 396

\bibitem[{{DasSarma} \& {DasSarma}(2018)}]{DD18}
{DasSarma}, P., \& {DasSarma}, S. 2018, Curr. Opin. Microbiol., 43, 24

\bibitem[{{de Duve}(2005)}]{deDu05}
{de Duve}, C. 2005, {Singularities: Landmarks on the Pathways of Life}
  (Cambridge University Press)

\bibitem[{{Deamer} \& {Weber}(2010)}]{DW10}
{Deamer}, D., \& {Weber}, A.~L. 2010, Cold Spring Harb. Perspect. Biol., 2,
  a004929

\bibitem[{{Deamer} \& {Oro}(1980)}]{DO80}
{Deamer}, D.~W., \& {Oro}, J. 1980, Biosystems, 12, 167

\bibitem[{{Delgado-Bonal}(2017)}]{DB17}
{Delgado-Bonal}, A. 2017, Sci. Rep., 7, 1642

\bibitem[{{Dieterich} {et~al.}(2012){Dieterich}, {Henry}, {Golimowski},
  {Krist}, \& {Tanner}}]{DHG12}
{Dieterich}, S.~B., {Henry}, T.~J., {Golimowski}, D.~A., {Krist}, J.~E., \&
  {Tanner}, A.~M. 2012, Astron. J., 144, 64

\bibitem[{{Dodd} {et~al.}(2017){Dodd}, {Papineau}, {Grenne}, {Slack},
  {Rittner}, {Pirajno}, {O'Neil}, \& {Little}}]{Dodd17}
{Dodd}, M.~S., {Papineau}, D., {Grenne}, T., {et~al.} 2017, Nature, 543, 60

\bibitem[{{Dole}(1964)}]{Dole64}
{Dole}, S.~H. 1964, {Habitable planets for man} (Blaisdell Pub.~Co.)

\bibitem[{{Donaldson} {et~al.}(2004){Donaldson}, {Tervahattu}, {Tuck}, \&
  {Vaida}}]{DTTV}
{Donaldson}, D.~J., {Tervahattu}, H., {Tuck}, A.~F., \& {Vaida}, V. 2004, Orig.
  Life Evol. Biosph., 34, 57

\bibitem[{{Donaldson} {et~al.}(2001){Donaldson}, {Tuck}, \& {Vaida}}]{DTV01}
{Donaldson}, D.~J., {Tuck}, A.~F., \& {Vaida}, V. 2001, Phys. Chem. Chem.
  Phys., 3, 5270

\bibitem[{{Dong} {et~al.}(2019){Dong}, {Huang}, \& {Lingam}}]{DHL19}
{Dong}, C., {Huang}, Z., \& {Lingam}, M. 2019, Astrophys. J. Lett., 882, L16

\bibitem[{{Dong} {et~al.}(2018){Dong}, {Jin}, {Lingam}, {Airapetian}, {Ma}, \&
  {van der Holst}}]{DJL18}
{Dong}, C., {Jin}, M., {Lingam}, M., {et~al.} 2018, Proc. Natl. Acad. Sci. USA,
  115, 260

\bibitem[{{Dong} {et~al.}(2017){Dong}, {Lingam}, {Ma}, \& {Cohen}}]{DLM17}
{Dong}, C., {Lingam}, M., {Ma}, Y., \& {Cohen}, O. 2017, Astrophys. J. Lett.,
  837, L26

\bibitem[{{Dyson}(1999)}]{Dys99}
{Dyson}, F. 1999, {Origins of Life}, 2nd edn. (Cambridge University Press)

\bibitem[{{Epps} {et~al.}(1979){Epps}, {Nooner}, {Eichberg}, {Sherwood}, \&
  {Or{\'o}}}]{ENE79}
{Epps}, D.~E., {Nooner}, D.~W., {Eichberg}, J., {Sherwood}, E., \& {Or{\'o}},
  J. 1979, J. Mol. Evol., 14, 235

\bibitem[{{Evans} {et~al.}(2012){Evans}, {Ireland}, {Kraus}, {Martinache},
  {Stewart}, {Tuthill}, {Lacour}, {Carpenter}, \& {Hillenbrand}}]{EIK12}
{Evans}, T.~M., {Ireland}, M.~J., {Kraus}, A.~L., {et~al.} 2012, Astrophys. J.,
  744, 120

\bibitem[{{Fegley} \& {Lodders}(1994)}]{FL94}
{Fegley}, Jr., B., \& {Lodders}, K. 1994, Icarus, 110, 117

\bibitem[{{Fontanive} {et~al.}(2018){Fontanive}, {Biller}, {Bonavita}, \&
  {Allers}}]{FBB18}
{Fontanive}, C., {Biller}, B., {Bonavita}, M., \& {Allers}, K. 2018, Mon. Not.
  R. Astron. Soc., 479, 2702

\bibitem[{{Forbes} \& {Loeb}(2019)}]{FL19}
{Forbes}, J.~C., \& {Loeb}, A. 2019, Astrophys. J., 871, 227

\bibitem[{{Freedman} {et~al.}(2014){Freedman}, {Lustig-Yaeger}, {Fortney},
  {Lupu}, {Marley}, \& {Lodders}}]{FLF14}
{Freedman}, R.~S., {Lustig-Yaeger}, J., {Fortney}, J.~J., {et~al.} 2014,
  Astrophys. J., Suppl. Ser., 214, 25

\bibitem[{{Fr{\"o}hlich-Nowoisky} {et~al.}(2016){Fr{\"o}hlich-Nowoisky},
  {Kampf}, {Weber}, {Huffman}, {P{\"o}hlker}, {Andreae}, {Lang-Yona},
  {Burrows}, {Gunthe}, {Elbert}, {Su}, {Hoor}, {Thines}, {Hoffmann},
  {Despr{\'e}s}, \& {P{\"o}schl}}]{FNK16}
{Fr{\"o}hlich-Nowoisky}, J., {Kampf}, C.~J., {Weber}, B., {et~al.} 2016,
  Atmospheric Res., 182, 346

\bibitem[{{Fujii} {et~al.}(2018){Fujii}, {Angerhausen}, {Deitrick},
  {Domagal-Goldman}, {Grenfell}, {Hori}, {Kane}, {Pall{\'e}}, {Rauer},
  {Siegler}, {Stapelfeldt}, \& {Stevenson}}]{FAD18}
{Fujii}, Y., {Angerhausen}, D., {Deitrick}, R., {et~al.} 2018, Astrobiology,
  18, 739

\bibitem[{{Ginsburg} {et~al.}(2018){Ginsburg}, {Lingam}, \& {Loeb}}]{GLL18}
{Ginsburg}, I., {Lingam}, M., \& {Loeb}, A. 2018, Astrophys. J. Lett., 868, L12

\bibitem[{{Goldacre}(1958)}]{Gold58}
{Goldacre}, R.~J. 1958, {Surface Films, their Collapse on Compression, the
  Shapes and Sizes of Cells and the Origin of Life}, ed. J.~F. {Danielli},
  K.~G.~A. {Parkhurst}, \& A.~C. {Riddiford} (Pergamon Press), 278--298

\bibitem[{{Goldblatt} \& {Watson}(2012)}]{GW12}
{Goldblatt}, C., \& {Watson}, A.~J. 2012, Phil. Trans. R. Soc. A, 370, 4197

\bibitem[{{Grether} \& {Lineweaver}(2006)}]{GL06}
{Grether}, D., \& {Lineweaver}, C.~H. 2006, Astrophys. J., 640, 1051

\bibitem[{{Griffith} {et~al.}(2012){Griffith}, {Tuck}, \& {Vaida}}]{GTV12}
{Griffith}, E.~C., {Tuck}, A.~F., \& {Vaida}, V. 2012, Acc. Chem. Res., 45,
  2106

\bibitem[{{Griffith} \& {Vaida}(2012)}]{GV12}
{Griffith}, E.~C., \& {Vaida}, V. 2012, Proc. Natl. Acad. Sci. USA, 109, 15697

\bibitem[{{Grinspoon}(1997)}]{Grin97}
{Grinspoon}, D.~H. 1997, {Venus Revealed: A New Look Below the Clouds of Our
  Mysterious Twin Planet} (Addison-Wesley)

\bibitem[{{Harrison} {et~al.}(2013){Harrison}, {Gheeraert}, {Tsigelnitskiy}, \&
  {Cockell}}]{HGT13}
{Harrison}, J.-P., {Gheeraert}, N., {Tsigelnitskiy}, D., \& {Cockell}, C.-S.
  2013, Trends Microbiol., 21, 204

\bibitem[{{Helling}(2019)}]{Hel19}
{Helling}, C. 2019, Annu Rev. Earth Planet. Sci., 47, 583

\bibitem[{{Helling} \& {Casewell}(2014)}]{HC14}
{Helling}, C., \& {Casewell}, S. 2014, Astron. Astrophys. Rev., 22, 80

\bibitem[{{Hill} \& {Rich}(1983)}]{HR83}
{Hill}, R., \& {Rich}, P.~R. 1983, Proc. Natl. Acad. Sci. USA, 80, 978

\bibitem[{{Hille}(2002)}]{Hil02}
{Hille}, R. 2002, Trends Biochem. Sci., 27, 360

\bibitem[{{Holl{\'o}sy}(2002)}]{Holl02}
{Holl{\'o}sy}, F. 2002, Micron, 33, 179

\bibitem[{{Horowitz} \& {Hubbard}(1974)}]{HH74}
{Horowitz}, N.~H., \& {Hubbard}, J.~S. 1974, Annu. Rev. Genetics, 8, 393

\bibitem[{{H{\"o}rst}(2017)}]{Hor17}
{H{\"o}rst}, S.~M. 2017, J. Geophys. Res. E, 122, 432

\bibitem[{{H{\"o}rst} \& {Tolbert}(2013)}]{HT13}
{H{\"o}rst}, S.~M., \& {Tolbert}, M.~A. 2013, Astrophys. J. Lett., 770, L10

\bibitem[{{Hubbard} {et~al.}(2002){Hubbard}, {Burrows}, \& {Lunine}}]{HBL02}
{Hubbard}, W.~B., {Burrows}, A., \& {Lunine}, J.~I. 2002, Annu. Rev. Astron.
  Astrophys., 40, 103

\bibitem[{{Jacob}(1999)}]{Jac99}
{Jacob}, D.~J. 1999, {Introduction to Atmospheric Chemistry} (Princeton
  University Press)

\bibitem[{{Kaltenegger}(2017)}]{Kal17}
{Kaltenegger}, L. 2017, Annu. Rev. Astron. Astrophys., 55, 433

\bibitem[{{Kasting} {et~al.}(1993){Kasting}, {Whitmire}, \& {Reynolds}}]{KWR93}
{Kasting}, J.~F., {Whitmire}, D.~P., \& {Reynolds}, R.~T. 1993, Icarus, 101,
  108

\bibitem[{{Keefe} \& {Miller}(1995)}]{KM95}
{Keefe}, A.~D., \& {Miller}, S.~L. 1995, J. Mol. Evol., 41, 693

\bibitem[{{Kiang} {et~al.}(2007{\natexlab{a}}){Kiang}, {Siefert}, {Govindjee},
  \& {Blankenship}}]{KSG07}
{Kiang}, N.~Y., {Siefert}, J., {Govindjee}, \& {Blankenship}, R.~E.
  2007{\natexlab{a}}, Astrobiology, 7, 222

\bibitem[{{Kiang} {et~al.}(2007{\natexlab{b}}){Kiang}, {Segura}, {Tinetti},
  {Govindjee}, {Blankenship}, {Cohen}, {Siefert}, {Crisp}, \&
  {Meadows}}]{KST07}
{Kiang}, N.~Y., {Segura}, A., {Tinetti}, G., {et~al.} 2007{\natexlab{b}},
  Astrobiology, 7, 252

\bibitem[{{Kirkpatrick} {et~al.}(2019){Kirkpatrick}, {Martin}, {Smart},
  {Cayago}, {Beichman}, {Marocco}, {Gelino}, {Faherty}, {Cushing}, {Schneider},
  {Mace}, {Tinney}, {Wright}, {Lowrance}, {Ingalls}, {Vrba}, {Munn}, {Dahm}, \&
  {McLean}}]{KMS19}
{Kirkpatrick}, J.~D., {Martin}, E.~C., {Smart}, R.~L., {et~al.} 2019,
  Astrophys. J., Suppl. Ser., 240, 19

\bibitem[{{Kitadai} \& {Maruyama}(2018)}]{KM18}
{Kitadai}, N., \& {Maruyama}, S. 2018, Geosci. Front., 9, 1117

\bibitem[{{Knoll}(2015)}]{Knoll15}
{Knoll}, A.~H. 2015, {Life on a Young Planet: The First Three Billion Years of
  Evolution on Earth}, Princeton Science Library (Princeton University Press)

\bibitem[{{Kopparapu} {et~al.}(2013){Kopparapu}, {Ramirez}, {Kasting}, {Eymet},
  {Robinson}, {Mahadevan}, {Terrien}, {Domagal-Goldman}, {Meadows}, \&
  {Deshpande}}]{KRK13}
{Kopparapu}, R.~K., {Ramirez}, R., {Kasting}, J.~F., {et~al.} 2013, Astrophys.
  J., 765, 131

\bibitem[{{Kowalchuk} \& {Stephen}(2001)}]{KS01}
{Kowalchuk}, G.~A., \& {Stephen}, J.~R. 2001, Annu. Rev. Microbiol., 55, 485

\bibitem[{{Kraus} \& {Hillenbrand}(2012)}]{KH12}
{Kraus}, A.~L., \& {Hillenbrand}, L.~A. 2012, Astrophys. J., 757, 141

\bibitem[{{Krissansen-Totton} {et~al.}(2016){Krissansen-Totton}, {Bergsman}, \&
  {Catling}}]{KTBC}
{Krissansen-Totton}, J., {Bergsman}, D.~S., \& {Catling}, D.~C. 2016,
  Astrobiology, 16, 39

\bibitem[{{Krissansen-Totton} {et~al.}(2018){Krissansen-Totton}, {Garland},
  {Irwin}, \& {Catling}}]{KTG18}
{Krissansen-Totton}, J., {Garland}, R., {Irwin}, P., \& {Catling}, D.~C. 2018,
  Astron. J., 156, 114

\bibitem[{{Kroupa} {et~al.}(2013){Kroupa}, {Weidner}, {Pflamm-Altenburg},
  {Thies}, {Dabringhausen}, {Marks}, \& {Maschberger}}]{KWP13}
{Kroupa}, P., {Weidner}, C., {Pflamm-Altenburg}, J., {et~al.} 2013, {The
  Stellar and Sub-Stellar Initial Mass Function of Simple and Composite
  Populations}, ed. T.~D. {Oswalt} \& G.~{Gilmore}, Vol.~5 (Springer), 115--242

\bibitem[{{Laakso} \& {Schrag}(2018)}]{LS18}
{Laakso}, T.~A., \& {Schrag}, D.~P. 2018, Global Biogeochem. Cy., 32, 486

\bibitem[{{Lafreni{\`e}re} {et~al.}(2007){Lafreni{\`e}re}, {Doyon}, {Marois},
  {Nadeau}, {Oppenheimer}, {Roche}, {Rigaut}, {Graham}, {Jayawardhana},
  {Johnstone}, {Kalas}, {Macintosh}, \& {Racine}}]{LD07}
{Lafreni{\`e}re}, D., {Doyon}, R., {Marois}, C., {et~al.} 2007, Astrophys. J.,
  670, 1367

\bibitem[{{Lammer} {et~al.}(2009){Lammer}, {Bredeh{\"o}ft}, {Coustenis},
  {Khodachenko}, {Kaltenegger}, {Grasset}, {Prieur}, {Raulin}, {Ehrenfreund},
  {Yamauchi}, {Wahlund}, {Grie{\ss}meier}, {Stangl}, {Cockell}, {Kulikov},
  {Grenfell}, \& {Rauer}}]{Lam09}
{Lammer}, H., {Bredeh{\"o}ft}, J.~H., {Coustenis}, A., {et~al.} 2009, Astron.
  Astrophys. Rev., 17, 181

\bibitem[{{Lide}(2007)}]{Lide07}
{Lide}, D.~R., ed. 2007, {CRC Handbook of Chemistry and Physics}, 88th edn.
  (CRC Press)

\bibitem[{{Limaye} {et~al.}(2018){Limaye}, {Mogul}, {Smith}, {Ansari},
  {S{\l}owik}, \& {Vaishampayan}}]{LMS18}
{Limaye}, S.~S., {Mogul}, R., {Smith}, D.~J., {et~al.} 2018, Astrobiology, 18,
  1181

\bibitem[{{Line} \& {Yung}(2013)}]{LY13}
{Line}, M.~R., \& {Yung}, Y.~L. 2013, Astrophys. J., 779, 3

\bibitem[{{Lingam}(2016)}]{Li16}
{Lingam}, M. 2016, Mon. Not. R. Astron. Soc., 455, 2792

\bibitem[{{Lingam} {et~al.}(2019){Lingam}, {Ginsburg}, \& {Bialy}}]{LGB19}
{Lingam}, M., {Ginsburg}, I., \& {Bialy}, S. 2019, Astrophys. J., 877, 62

\bibitem[{{Lingam} \& {Loeb}(2017)}]{LL17}
{Lingam}, M., \& {Loeb}, A. 2017, Astrophys. J. Lett., 846, L21

\bibitem[{{Lingam} \& {Loeb}(2018{\natexlab{a}})}]{Lin18}
---. 2018{\natexlab{a}}, Int. J. Astrobiol., 17, 116

\bibitem[{{Lingam} \& {Loeb}(2018{\natexlab{b}})}]{LL18}
---. 2018{\natexlab{b}}, Astron. J., 156, 151

\bibitem[{{Lingam} \& {Loeb}(2019{\natexlab{a}})}]{LL19}
---. 2019{\natexlab{a}}, Int. J. Astrobiol., doi:10.1017/S1473550419000016

\bibitem[{{Lingam} \& {Loeb}(2019{\natexlab{b}})}]{Man19}
---. 2019{\natexlab{b}}, Rev. Mod. Phys., 91, 021002

\bibitem[{{Lingam} \& {Loeb}(2019{\natexlab{c}})}]{LiLo19}
---. 2019{\natexlab{c}}, Int. J. Astrobiol., 18, 112

\bibitem[{{Lingam} \& {Loeb}(2019{\natexlab{d}})}]{MaLi19}
---. 2019{\natexlab{d}}, Astron. J., 157, 25

\bibitem[{{Lingam} \& {Loeb}(2019{\natexlab{e}})}]{Ling19}
---. 2019{\natexlab{e}}, Mon. Not. R. Astron. Soc., 485, 5924

\bibitem[{{Lodders} \& {Fegley}(2002)}]{LF02}
{Lodders}, K., \& {Fegley}, B. 2002, Icarus, 155, 393

\bibitem[{{Lohrmann} \& {Orgel}(1971)}]{LO71}
{Lohrmann}, R., \& {Orgel}, L.~E. 1971, Science, 171, 490

\bibitem[{{Luger} \& {Barnes}(2015)}]{LB15}
{Luger}, R., \& {Barnes}, R. 2015, Astrobiology, 15, 119

\bibitem[{{Luhman}(2014)}]{Luh14}
{Luhman}, K.~L. 2014, Astrophys. J. Lett., 786, L18

\bibitem[{{Luisi}(2016)}]{Lu16}
{Luisi}, P.~L. 2016, The Emergence of Life: From Chemical Origins to Synthetic
  Biology (Cambridge Univ. Press)

\bibitem[{{Lunine}(2017)}]{Lun17}
{Lunine}, J.~I. 2017, Acta Astronaut., 131, 123

\bibitem[{{Ma} \& {Ge}(2014)}]{MG14}
{Ma}, B., \& {Ge}, J. 2014, Mon. Not. R. Astron. Soc., 439, 2781

\bibitem[{{Ma} {et~al.}(2018){Ma}, {He}, {Tian}, {Zou}, {Yan}, {Yang}, {Zhou},
  {Huang}, {Shen}, \& {Fang}}]{MHT18}
{Ma}, S., {He}, F., {Tian}, D., {et~al.} 2018, Biogeosciences, 15, 693

\bibitem[{{Marcy} \& {Butler}(2000)}]{MB00}
{Marcy}, G.~W., \& {Butler}, R.~P. 2000, Publ. Astron. Soc. Pac., 112, 137

\bibitem[{{Marley} \& {Robinson}(2015)}]{MR15}
{Marley}, M.~S., \& {Robinson}, T.~D. 2015, Annu. Rev. Astron. Astrophys., 53,
  279

\bibitem[{{McCann}(1968)}]{McC68}
{McCann}, H.~G. 1968, Arch. Oral Biol., 13, 987

\bibitem[{{McKay}(2014)}]{McK14}
{McKay}, C.~P. 2014, Proc. Natl. Acad. Sci. USA, 111, 12628

\bibitem[{{Metchev} \& {Hillenbrand}(2009)}]{MH09}
{Metchev}, S.~A., \& {Hillenbrand}, L.~A. 2009, Astrophys. J., Suppl. Ser.,
  181, 62

\bibitem[{{Morley} {et~al.}(2014){Morley}, {Marley}, {Fortney}, {Lupu},
  {Saumon}, {Greene}, \& {Lodders}}]{MMF14}
{Morley}, C.~V., {Marley}, M.~S., {Fortney}, J.~J., {et~al.} 2014, Astrophys.
  J., 787, 78

\bibitem[{{Morley} {et~al.}(2018){Morley}, {Skemer}, {Allers}, {Marley},
  {Faherty}, {Visscher}, {Beiler}, {Miles}, {Lupu}, {Freedman}, {Fortney},
  {Geballe}, \& {Bjoraker}}]{MSA18}
{Morley}, C.~V., {Skemer}, A.~J., {Allers}, K.~N., {et~al.} 2018, Astrophys.
  J., 858, 97

\bibitem[{{Morowitz} \& {Sagan}(1967)}]{MS67}
{Morowitz}, H., \& {Sagan}, C. 1967, Nature, 215, 1259

\bibitem[{{Nimmo} \& {Pappalardo}(2016)}]{NP16}
{Nimmo}, F., \& {Pappalardo}, R.~T. 2016, . Geophys. Res. E, 121, 1378

\bibitem[{{Nordstrom} \& {Southam}(1997)}]{NS97}
{Nordstrom}, D.~K., \& {Southam}, G. 1997, Rev. Mineralogy, 35, 381

\bibitem[{{N{\"u}rnberg} {et~al.}(2018){N{\"u}rnberg}, {Morton},
  {Santabarbara}, {Telfer}, {Joliot}, {Antonaru}, {Ruban}, {Cardona}, {Krausz},
  {Boussac}, {Fantuzzi}, \& {Rutherford}}]{NMS18}
{N{\"u}rnberg}, D.~J., {Morton}, J., {Santabarbara}, S., {et~al.} 2018,
  Science, 360, 1210

\bibitem[{{Or{\'o}} \& {Stephen-Sherwood}(1976)}]{OSS76}
{Or{\'o}}, J., \& {Stephen-Sherwood}, E. 1976, Orig. Life, 7, 37

\bibitem[{{Pasek} {et~al.}(2017){Pasek}, {Gull}, \& {Herschy}}]{PGH17}
{Pasek}, M.~A., {Gull}, M., \& {Herschy}, B. 2017, Chem. Geol., 475, 149

\bibitem[{{Ponnamperuma}(1976)}]{Pon76}
{Ponnamperuma}, C. 1976, Icarus, 29, 321

\bibitem[{{Ponnamperuma} \& {Mack}(1965)}]{PM65}
{Ponnamperuma}, C., \& {Mack}, R. 1965, Science, 148, 1221

\bibitem[{{Ponnamperuma} \& {Molton}(1973)}]{PM73}
{Ponnamperuma}, C., \& {Molton}, P. 1973, Space Life Sci., 4, 32

\bibitem[{{Powner} \& {Sutherland}(2011)}]{PS11}
{Powner}, M.~W., \& {Sutherland}, J.~D. 2011, Phil. Trans. R. Soc. B, 366, 2870

\bibitem[{{Preston} \& {Dartnell}(2014)}]{PD14}
{Preston}, L.~J., \& {Dartnell}, L.~R. 2014, Int. J. Astrobiol., 13, 81

\bibitem[{{Radigan} {et~al.}(2012){Radigan}, {Jayawardhana}, {Lafreni{\`e}re},
  {Artigau}, {Marley}, \& {Saumon}}]{RJL12}
{Radigan}, J., {Jayawardhana}, R., {Lafreni{\`e}re}, D., {et~al.} 2012,
  Astrophys. J., 750, 105

\bibitem[{{Ramirez}(2018)}]{Ram18}
{Ramirez}, R.~M. 2018, Geosciences, 8, 280

\bibitem[{{Raven} {et~al.}(2000){Raven}, {K{\"u}bler}, \& {Beardall}}]{RKB00}
{Raven}, J.~A., {K{\"u}bler}, J.~E., \& {Beardall}, J. 2000, J. Mar. Biol.
  Assoc. UK, 80, 1

\bibitem[{{Reggiani} {et~al.}(2016){Reggiani}, {Meyer}, {Chauvin}, {Vigan},
  {Quanz}, {Biller}, {Bonavita}, {Desidera}, {Delorme}, {Hagelberg}, {Maire},
  {Boccaletti}, {Beuzit}, {Buenzli}, {Carson}, {Covino}, {Feldt}, {Girard},
  {Gratton}, {Henning}, {Kasper}, {Lagrange}, {Mesa}, {Messina}, {Montagnier},
  {Mordasini}, {Mouillet}, {Schlieder}, {Segransan}, {Thalmann}, \&
  {Zurlo}}]{RMC16}
{Reggiani}, M., {Meyer}, M.~R., {Chauvin}, G., {et~al.} 2016, Astron.
  Astrophys., 586, A147

\bibitem[{{Ritchie} {et~al.}(2018){Ritchie}, {Larkum}, \& {Ribas}}]{RLR18}
{Ritchie}, R.~J., {Larkum}, A.~W.~D., \& {Ribas}, I. 2018, Int. J. Astrobiol.,
  17, 147

\bibitem[{{Rushby} {et~al.}(2013){Rushby}, {Claire}, {Osborn}, \&
  {Watson}}]{RCO13}
{Rushby}, A.~J., {Claire}, M.~W., {Osborn}, H., \& {Watson}, A.~J. 2013,
  Astrobiology, 13, 833

\bibitem[{{Ryde} {et~al.}(2002){Ryde}, {Lambert}, {Richter}, \& {Lacy}}]{RLR02}
{Ryde}, N., {Lambert}, D.~L., {Richter}, M.~J., \& {Lacy}, J.~H. 2002,
  Astrophys. J., 580, 447

\bibitem[{{Sagan}(1960)}]{Sag60}
{Sagan}, C. 1960, Astron. J., 65, 499

\bibitem[{{Sagan} \& {Salpeter}(1976)}]{SS76}
{Sagan}, C., \& {Salpeter}, E.~E. 1976, Astrophys. J., Suppl. Ser., 32, 737

\bibitem[{{Sahlmann} {et~al.}(2011){Sahlmann}, {S{\'e}gransan}, {Queloz},
  {Udry}, {Santos}, {Marmier}, {Mayor}, {Naef}, {Pepe}, \& {Zucker}}]{SSQ11}
{Sahlmann}, J., {S{\'e}gransan}, D., {Queloz}, D., {et~al.} 2011, Astron.
  Astrophys., 525, A95

\bibitem[{{Sarmiento} \& {Gruber}(2006)}]{SG06}
{Sarmiento}, J.~L., \& {Gruber}, N. 2006, {Ocean Biogeochemical Dynamics}
  (Princeton University Press)

\bibitem[{{Scharf}(2019)}]{Sch19}
{Scharf}, C. 2019, Astrophys. J., 876, 16

\bibitem[{{Schulze-Makuch} {et~al.}(2004){Schulze-Makuch}, {Grinspoon},
  {Abbas}, {Irwin}, \& {Bullock}}]{SGA04}
{Schulze-Makuch}, D., {Grinspoon}, D.~H., {Abbas}, O., {Irwin}, L.~N., \&
  {Bullock}, M.~A. 2004, Astrobiology, 4, 11

\bibitem[{{Schulze-Makuch} \& {Irwin}(2008)}]{SMI08}
{Schulze-Makuch}, D., \& {Irwin}, L.~N. 2008, {Life in the Universe}, 2nd edn.
  (Springer-Verlag), doi:10.1007/978-3-540-76817-3

\bibitem[{{Schwarz} {et~al.}(2009){Schwarz}, {Mendel}, \& {Ribbe}}]{SMR09}
{Schwarz}, G., {Mendel}, R.~R., \& {Ribbe}, M.~W. 2009, Nature, 460, 839

\bibitem[{{Schwieterman} {et~al.}(2018){Schwieterman}, {Kiang}, {Parenteau},
  {Harman}, {DasSarma}, {Fisher}, {Arney}, {Hartnett}, {Reinhard}, {Olson},
  {Meadows}, {Cockell}, {Walker}, {Grenfell}, {Hegde}, {Rugheimer}, {Hu}, \&
  {Lyons}}]{SKP18}
{Schwieterman}, E.~W., {Kiang}, N.~Y., {Parenteau}, M.~N., {et~al.} 2018,
  Astrobiology, 18, 663

\bibitem[{{Seager} {et~al.}(2005){Seager}, {Turner}, {Schafer}, \&
  {Ford}}]{STSF}
{Seager}, S., {Turner}, E.~L., {Schafer}, J., \& {Ford}, E.~B. 2005,
  Astrobiology, 5, 372

\bibitem[{{Seckbach} \& {Libby}(1970)}]{SL70}
{Seckbach}, J., \& {Libby}, W.~F. 1970, Space Life Sci., 2, 121

\bibitem[{{Shapley}(1967)}]{Shap67}
{Shapley}, H. 1967, {Beyond the Observatory} (Charles Scribner's Sons)

\bibitem[{{Shields} {et~al.}(2016){Shields}, {Ballard}, \& {Johnson}}]{SBJ16}
{Shields}, A.~L., {Ballard}, S., \& {Johnson}, J.~A. 2016, Phys. Rep., 663, 1

\bibitem[{{Showman} \& {Kaspi}(2013)}]{SK13}
{Showman}, A.~P., \& {Kaspi}, Y. 2013, Astrophys. J., 776, 85

\bibitem[{{Skemer} {et~al.}(2016){Skemer}, {Morley}, {Allers}, {Geballe},
  {Marley}, {Fortney}, {Faherty}, {Bjoraker}, \& {Lupu}}]{SMA16}
{Skemer}, A.~J., {Morley}, C.~V., {Allers}, K.~N., {et~al.} 2016, Astrophys. J.
  Lett., 826, L17

\bibitem[{{Smith}(2013)}]{Smi13}
{Smith}, D.~J. 2013, Astrobiology, 13, 981

\bibitem[{{Smith} {et~al.}(2010){Smith}, {Griffin}, \& {Schuerger}}]{SGS10}
{Smith}, D.~J., {Griffin}, D.~W., \& {Schuerger}, A.~C. 2010, Aerobiologia, 26,
  35

\bibitem[{{Sodek} \& {Redmond}(1967)}]{SoRe67}
{Sodek}, B.~A., \& {Redmond}, J.~C. 1967, BioScience, 17, 97

\bibitem[{{Spohn} \& {Schubert}(2003)}]{SS03}
{Spohn}, T., \& {Schubert}, G. 2003, Icarus, 161, 456

\bibitem[{{Stark} {et~al.}(2014){Stark}, {Helling}, {Diver}, \&
  {Rimmer}}]{SHD14}
{Stark}, C.~R., {Helling}, C., {Diver}, D.~A., \& {Rimmer}, P.~B. 2014, Int. J.
  Astrobiol., 13, 165

\bibitem[{{Stribling} \& {Miller}(1987)}]{SM87}
{Stribling}, R., \& {Miller}, S.~L. 1987, Icarus, 72, 48

\bibitem[{{Thies} \& {Kroupa}(2007)}]{TK07}
{Thies}, I., \& {Kroupa}, P. 2007, Astrophys. J., 671, 767

\bibitem[{{Thies} {et~al.}(2015){Thies}, {Pflamm-Altenburg}, {Kroupa}, \&
  {Marks}}]{TPK15}
{Thies}, I., {Pflamm-Altenburg}, J., {Kroupa}, P., \& {Marks}, M. 2015,
  Astrophys. J., 800, 72

\bibitem[{{Tomasko} {et~al.}(2008){Tomasko}, {Doose}, {Engel}, {Dafoe}, {West},
  {Lemmon}, {Karkoschka}, \& {See}}]{TDE08}
{Tomasko}, M.~G., {Doose}, L., {Engel}, S., {et~al.} 2008, Planet. Space Sci.,
  56, 669

\bibitem[{{Tuck}(2002)}]{Tuck02}
{Tuck}, A. 2002, Surv. Geophys., 23, 379

\bibitem[{{Tyrrell}(1999)}]{Tyrr99}
{Tyrrell}, T. 1999, Nature, 400, 525

\bibitem[{{Visscher} {et~al.}(2006){Visscher}, {Lodders}, \& {Fegley}}]{VLF06}
{Visscher}, C., {Lodders}, K., \& {Fegley}, Jr., B. 2006, Astrophys. J., 648,
  1181

\bibitem[{{Visscher} {et~al.}(2010){Visscher}, {Lodders}, \& {Fegley}}]{VLF10}
---. 2010, Astrophys. J., 716, 1060

\bibitem[{{Vladilo} \& {Hassanali}(2018)}]{VH18}
{Vladilo}, G., \& {Hassanali}, A. 2018, Life, 8, 1

\bibitem[{{Wachtershauser}(1990)}]{Wac90}
{Wachtershauser}, G. 1990, Proc. Natl. Acad. Sci. USA, 87, 200

\bibitem[{{Wainwright} {et~al.}(2006){Wainwright}, {Alharbi}, \&
  {Wickramasinghe}}]{WAW06}
{Wainwright}, M., {Alharbi}, S., \& {Wickramasinghe}, N.~C. 2006, Int. J.
  Astrobiol., 5, 13

\bibitem[{{West} {et~al.}(2004){West}, {Baines}, {Friedson}, {Banfield},
  {Ragent}, \& {Taylor}}]{WBF04}
{West}, R.~A., {Baines}, K.~H., {Friedson}, A.~J., {et~al.} 2004, {Jovian
  clouds and haze}, ed. F.~{Bagenal}, T.~E. {Dowling}, \& W.~B. {McKinnon},
  Vol.~1 (Cambridge University Press), 79--104

\bibitem[{{West} {et~al.}(2009){West}, {Baines}, {Karkoschka}, \&
  {S{\'a}nchez-Lavega}}]{WBK09}
{West}, R.~A., {Baines}, K.~H., {Karkoschka}, E., \& {S{\'a}nchez-Lavega}, A.
  2009, {Clouds and Aerosols in Saturn's Atmosphere}, ed. M.~K. {Dougherty},
  L.~W. {Esposito}, \& S.~M. {Krimigis} (Springer), 161--179

\bibitem[{{Westheimer}(1987)}]{West87}
{Westheimer}, F.~H. 1987, Science, 235, 1173

\bibitem[{{Wilson} {et~al.}(2016){Wilson}, {H{\'e}brard}, {Santos}, {Sahlmann},
  {Montagnier}, {Astudillo-Defru}, {Boisse}, {Bouchy}, {Rey}, {Arnold},
  {Bonfils}, {Bourrier}, {Courcol}, {Deleuil}, {Delfosse}, {D{\'{\i}}az},
  {Ehrenreich}, {Forveille}, {Moutou}, {Pepe}, {Santerne}, {S{\'e}gransan}, \&
  {Udry}}]{WHS16}
{Wilson}, P.~A., {H{\'e}brard}, G., {Santos}, N.~C., {et~al.} 2016, Astron.
  Astrophys., 588, A144

\bibitem[{{Winn}(2010)}]{Winn10}
{Winn}, J.~N. 2010, {Exoplanet Transits and Occultations}, ed. S.~{Seager}
  (University of Arizona Press), 55--77

\bibitem[{{Wolstencroft} \& {Raven}(2002)}]{WoRa02}
{Wolstencroft}, R.~D., \& {Raven}, J.~A. 2002, Icarus, 157, 535

\bibitem[{{Yates} {et~al.}(2017){Yates}, {Palmer}, {Biller}, \&
  {Cockell}}]{YPB17}
{Yates}, J.~S., {Palmer}, P.~I., {Biller}, B., \& {Cockell}, C.~S. 2017,
  Astrophys. J., 836, 184

\bibitem[{{Zahnle} \& {Marley}(2014)}]{ZM14}
{Zahnle}, K.~J., \& {Marley}, M.~S. 2014, Astrophys. J., 797, 41

\bibitem[{{Zalesky} {et~al.}(2019){Zalesky}, {Line}, {Schneider}, \&
  {Patience}}]{ZLSP19}
{Zalesky}, J.~A., {Line}, M.~R., {Schneider}, A.~C., \& {Patience}, J. 2019,
  Astrophys. J., 877, 24

\bibitem[{{Zasova} {et~al.}(1981){Zasova}, {Krasnopolskii}, \& {Moroz}}]{ZKM81}
{Zasova}, L.~V., {Krasnopolskii}, V.~A., \& {Moroz}, V.~I. 1981, Adv. Space
  Res., 1, 13

\bibitem[{{Zhang} \& {Showman}(2014)}]{ZS14}
{Zhang}, X., \& {Showman}, A.~P. 2014, Astrophys. J. Lett., 788, L6

\bibitem[{{Zink} \& {Hansen}(2019)}]{ZH19}
{Zink}, J.~K., \& {Hansen}, B.~M.~S. 2019, Mon. Not. R. Astron. Soc., 487, 246

\end{thebibliography}

\end{document}